\documentclass[conference]{IEEEtran}
\IEEEoverridecommandlockouts
\usepackage{cite}
\usepackage{amsmath,amssymb,amsfonts}
\usepackage{graphicx}
\usepackage{textcomp}
\usepackage{xcolor}

\usepackage{booktabs}
\usepackage[misc]{ifsym}
\usepackage[linesnumbered,longend,ruled,vlined,boxed,commentsnumbered]{algorithm2e}
\usepackage{algpseudocode}
\usepackage{multirow}
\usepackage{subcaption} 
\usepackage{url}
\usepackage{ragged2e} 
\usepackage{colortbl}
\usepackage[table]{xcolor} 
\SetAlgoBlockMarkers{end}{}
\definecolor{ncncolor}{HTML}{4F6272}
\definecolor{ncnccolor}{HTML}{E1B16A}
\definecolor{cglenchcolor}{HTML}{D48C6A}
\definecolor{cglencnccolor}{HTML}{6A9387}
\definecolor{deepgreen}{rgb}{0.0, 0.5, 0.0}

\def\BibTeX{{\rm B\kern-.05em{\sc i\kern-.025em b}\kern-.08em
    T\kern-.1667em\lower.7ex\hbox{E}\kern-.125emX}}
    
\begin{document}

% User-defined theorem environments
\newtheorem{prop}{Proposition}
\newtheorem{theorem}{Theorem}

% User-defined spacing adjustments
\textfloatsep 1.1ex plus.1ex
\intextsep 1.1ex plus.1ex
\floatsep 1.1ex plus.21ex

\title{CGLE: Class-label Graph Link Estimator for Link Prediction\\
}

\author{\IEEEauthorblockN{Ankit Mazumder}
\IEEEauthorblockA{\textit{Yardi School of Artificial Intelligence} \\
\textit{Indian Institute of Technology, Delhi}\\
Hauz Khas, New Delhi, Delhi, India \\
aiy227513@scai.iitd.ac.in}

\and
\IEEEauthorblockN{Srikanta Bedathur}
\IEEEauthorblockA{\textit{Yardi School of Artificial Intelligence} \\
\textit{Department of Computer Science and Engineering} \\
\textit{Indian Institute of Technology, Delhi}\\
Hauz Khas, New Delhi, Delhi, India \\
srikanta@cse.iitd.ac.in}

% \and
% \IEEEauthorblockN{3\textsuperscript{rd} Author Name}
% \IEEEauthorblockA{\textit{dept. name of organization} \\
% \textit{name of organization}\\
}

\maketitle

\begin{abstract}
Link prediction is a pivotal task in graph mining with wide-ranging applications in social networks, recommendation systems, and knowledge graph completion. However, many leading Graph Neural Network (GNN) models often neglect the valuable semantic information aggregated at the class level. To address this limitation, this paper introduces \textbf{CGLE (Class-label Graph Link Estimator)}, a novel framework designed to augment GNN-based link prediction models. CGLE operates by constructing a class-conditioned link probability matrix, where each entry represents the probability of a link forming between two node classes. This matrix is derived from either available ground-truth labels or from pseudo-labels obtained through clustering. The resulting class-based prior is then concatenated with the structural link embedding from a backbone GNN, and the combined representation is processed by a Multi-Layer Perceptron (MLP) for the final prediction. Crucially, CGLE's logic is encapsulated in an efficient preprocessing stage, leaving the computational complexity of the underlying GNN model unaffected. We validate our approach through extensive experiments on a broad suite of benchmark datasets, covering both homophilous and sparse heterophilous graphs. The results show that CGLE yields substantial performance gains over strong baselines like NCN/NCNC, with improvements in \textbf{HR@100 of over 10\% points on homophilous datasets like Pubmed and DBLP}. On the sparse heterophilous graphs, CGLE delivers an \textbf{MRR improvement of over 4\% on the Chameleon dataset}. Our work underscores the efficacy of integrating global, data-driven semantic priors, presenting a compelling alternative to the pursuit of ever-more-complex model architectures. Code to reproduce our findings is available at \url{https://github.com/data-iitd/cgle-icdm2025}.
\end{abstract}

\section{Introduction}

Graphs serve as a fundamental representation for complex systems across diverse domains, including social networks, biological interactions, knowledge graphs, and recommendation systems~\cite{DBLP:books/daglib/0025903,Liben-Nowell:2003,10.1145/3543507.3583340,DBLP:conf/www/LeskovecHK10,Martnez2016ASO,Kovcs2018NetworkbasedPO,Yue2019GraphEO}. A key challenge in graph-based learning is \textit{link prediction (LP)}, which aims to estimate the likelihood of missing or future edges between nodes~\cite{Liben-Nowell:2003,zhang2018link,Martnez2016ASO,DBLP:conf/kdd/TylendaAB09}. Accurate link prediction plays a crucial role in applications such as recommendation systems, biological discovery, and fraud detection.

\begin{figure}[t]
    \centering
    \includegraphics[width=0.8\linewidth]{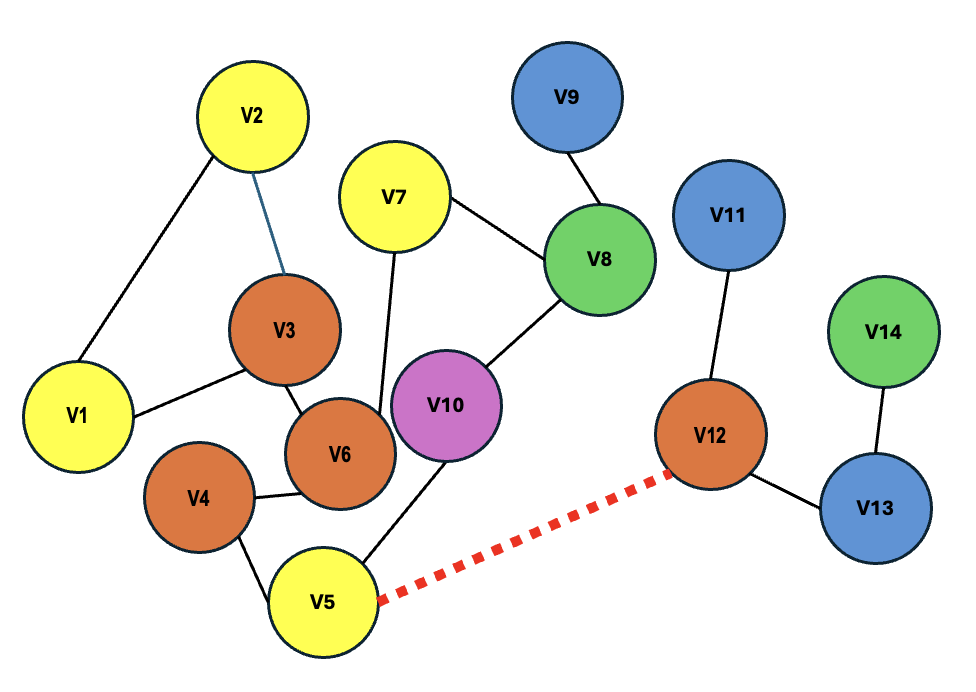}
    \caption{In this illustration of class-label guided link prediction, node colors represent their class. The goal is to predict a link between the disconnected nodes V5 and V12. Standard methods like CN, which underpin top models, would fail to predict this link. CGLE, however, can predict the connection by incorporating the nodes' class information and identifying the significant global co-occurrence pattern between the yellow and orange classes.}
    \label{fig:link_prediction_example}
\end{figure}

Early link prediction techniques used hand-crafted heuristics on the structural properties of the graph such as common-neighbours and Adamic-Adar index~\cite{Liben-Nowell:2003,Adamic:2003}. With the advent of graph neural networks (GNNs), they have been used successfully in several graph learning tasks, including link prediction~\cite{li2023evaluatinggraphneuralnetworks}. GNNs generate expressive node embeddings that capture both the local and global structural properties, as well as additional features on the nodes/edges that were typically ignored by earlier methods~\cite{gilmer2017neural,DBLP:conf/iclr/KipfW17,Hamilton:2017tp,Velickovic:2018we}. Despite their success for tasks such as node / graph classification, GNN-based models often predict links as a function of embeddings of the nodes involved in the link being predicted. This node-centric paradigm of link prediction fails to capture pairwise relationships among nodes effectively, resulting in suboptimal performance on heterophilic and structurally complex networks. There have been several increasingly sophisticated attempts to overcome this limitation of link prediction by various structural features of graphs surrounding the node pairs~\cite{Chamberlain2022GraphNN,yun2021neognns,ncnc2024,DBLP:conf/aaai/ShiHZHZZ24}. 

In this paper, we propose a conceptually simple novel GNN-based link prediction model called CGLE (Class-label Guided Link Estimator) which enhances the link prediction performance by exploiting the aggregate properties of node pairs involved in the link prediction beyond just their structural neighborhoods. Specifically, CGLE takes advantage of \textit{class labels} in the graph, or inferred clusters when these class labels are not readily available, to model global link formation priors, which can then be incorporated into any base GNN model. The primary advantage of this approach is that it breaks out of the dependence on structural neighborhood alone, since the class labels or clusterings of nodes can incorporate the node features as well as their topological features. 

Our class-label guided model, CGLE, overcomes the limitations of conventional heuristics by capturing global co-occurrence patterns between node classes. For example, in Fig.~\ref{fig:link_prediction_example}, CGLE correctly identifies the link between \(V_5\) and \(V_{12}\) using a probability matrix \(P\), whose entries are defined in Equation~\ref{eq:equation7} and visualized in Fig.~\ref{fig:heatmap_analysis_2x3}. Our implementation (Fig.~\ref{fig:cgle_arch}) is based on the recent NCN and NCNC models~\cite{ncnc2024} and is widely adaptable. Extensive evaluations on various graph structures confirm that CGLE consistently improves link prediction performance by 2-6\% over the base models.

In summary, our contributions are as follows: 
\begin{enumerate}
\item We introduce CGLE, a conceptually simple, highly adaptable, and computationally efficient model that utilizes aggregate properties of node pairs to boost the link prediction performance of base GNN models. 
\item We show that CGLE can be used effectively with a variety of datasets both, homophilous and heterophilous; graphs with pre-defined node-level labels and those where we independently derive cluster labels. 
\item We empirically demonstrate the clear and remarkable performance gains of CGLE across benchmarks over base GNN models.
\end{enumerate}

\subsection{Organization}
The remainder of this paper is organized as follows. Section~\ref{sec: related_work} reviews related work, and Section~\ref{sec: theory} introduces the theoretical foundations of CGLE. We detail our implementation and model architecture in Section~\ref{sec: implemtation}, followed by the experimental setup and datasets in Section~\ref{sec: experimentation}. Section~\ref{sec: psudo_class} explains the process for generating pseudo-class labels for datasets that lack them. In Section~\ref{sec:ablation}, we present ablation studies on alternative clustering methods (Louvain and spectral) and the impact of mono-labeling. Section~\ref{sec: challenges} outlines key challenges and future directions, and Section~\ref{sec: conclude} concludes the paper.

\section{Related Work}
\label{sec: related_work}
Due to its importance, there is a large body of work on link prediction models for graphs. Early models used various topological characteristics including the shortest distance in the graph, common neighbors, preferential attachment, Adamic-Adar~\cite{Liben-Nowell:2003,Adamic:2003}, Jaccard~\cite{6268901}, SimRank~\cite{jeh2002simrank}, rooted PageRank~\cite{brin1998anatomy}, and Katz index~\cite{katz1953new}, to predict the probability of link formation~\cite{Liben-Nowell:2003,Barabasi:1999}. Embedding-based approaches, like Matrix Factorization (MF)~\cite{koren2009matrix}, Multilayer Perceptron (MLP)~\cite{rumelhart1986learning}, and Node2Vec~\cite{Grover:2016}, learn node embeddings for link prediction. A major shortcoming of these approaches is their use of only the topological features for link prediction task. 

GNN-based methods, such as Graph Convolutional Networks (GCN)~\cite{DBLP:conf/iclr/KipfW17}, Graph Attention Networks (GAT)~\cite{Velickovic:2018we}, and GraphSAGE~\cite{Hamilton:2017tp}, adopt message passing to aggregate information from node neighborhoods across multiple hops. These methods effectively combine node features with structural information, achieving superior performance.

Recent models have advanced link prediction by incorporating richer structural and positional information. For instance, some approaches focus on local topology, with SEAL~\cite{zhang2018link} extracting k-hop subgraphs and models like Neo-GNN~\cite{yun2021neognns}, NCN, and NCNC~\cite{ncnc2024} integrating common neighbor information. Others leverage different architectural enhancements, such as PEG~\cite{wang2022equivariant}, which uses positional encoding for improved relational modeling. This trend of creating hybrid models that integrate pairwise features with graph structure is shared by other state-of-the-art methods like NBFnet~\cite{Zhu2021NeuralBN} and BUDDY~\cite{chamberlain2023graph}.

There is also previous work on the use of higher-order structural features for link prediction~\cite{DBLP:journals/pnas/BensonASJK18,DBLP:conf/pkdd/AbuodaMA19,Battiston2020NetworksBP}, based on the insight that link formation is influenced not only by the pairwise node features, but also by the higher-order substructure within the graph they are embedded in. CGLE is also based on a similar insight, but goes beyond the use of just topology-driven substructures. Instead, it utilizes the node class labels, which clearly define a semantic grouping of nodes, or derived labels for nodes based on their clustering based on structural as well as node-level features. 

\section{Class-Label-Based Link Prediction Theory}
\label{sec: theory}

Existing link prediction heuristics primarily leverage structural patterns within local subgraphs~\cite{zhang2018link,yun2021neognns,Zhu2021NeuralBN,Chamberlain2022GraphNN,ncnc2024}, but they often overlook the role of node class distributions. To address this limitation, we extend the theoretical framework by formally integrating class labels into the link prediction process.

\subsection{Baseline Theorem: $\gamma$-Decaying Structural Heuristic~\cite{zhang2018link}}

\begin{theorem}
Let \( x, y \in V \) be a pair of nodes in an undirected graph \( G = (V, E) \). A $\gamma$-decaying structural heuristic for link prediction between \( x \) and \( y \) is defined as:
\begin{equation}
    H(x, y) = \eta \sum_{l=1}^{\infty} \gamma^l f(x, y, l),
\end{equation}
where \( \gamma \in (0,1) \) is a decay factor, \( \eta > 0 \) is a bounded scaling constant, and \( f(x, y, l) \) encodes structural features (e.g., number of walks or path-based statistics) of length \( l \) between nodes \( x \) and \( y \).

If \( f(x, y, l) \leq \lambda^l \) for some \( \lambda < \frac{1}{\gamma} \), and \( f(x, y, l) \) is computable from the \( h \)-hop enclosing subgraph \( G^h_{x,y} \) for all \( l \leq g(h) \), with \( g(h) = ah + b \) for constants \( a > 0 \), \( b \in \mathbb{N} \), then the heuristic \( H(x, y) \) can be approximated using only \( G^h_{x,y} \), with approximation error decreasing at least exponentially with \( h \).
\end{theorem}

\vspace{0.5em}
This result provides a theoretical foundation for learning link prediction heuristics from local subgraphs. It unifies classical heuristics such as Katz index~\cite{katz1953new}, rooted PageRank~\cite{brin1998anatomy}, and SimRank~\cite{jeh2002simrank} under a common framework, showing that they can be effectively approximated using localized graph structure without requiring access to the global network.

We encourage the readers to refer~\cite{zhang2018link} for the proof and further theoretical insights of the above theorem. We now present a refinement of the theorem which incorporates the class-label probability in link prediction.

\subsection{{Refined Proposition: Class-Label Probability in Link Prediction}}

\noindent\textbf{Intuition and Motivation.} Traditional structural heuristics, such as the Katz index or rooted PageRank, effectively model proximity-based link patterns but often fall short in sparse or heterophilous graphs where structural cues alone may be insufficient. As illustrated in Fig.~\ref{fig:link_prediction_example}, incorporating class-label information significantly enhances link prediction by:

\begin{itemize}
\item \textit{Improving Disambiguation.} Class labels provide additional node-specific attributes, distinguishing structurally similar nodes.
\item \textit{Capturing Long-Range Dependencies.} Class-label dependencies extend beyond local neighborhoods, enriching the model with non-local semantic information.
\item \textit{Mitigating Data Sparsity.} Class-based similarities reveal latent connections that may be overlooked by purely structural features, particularly in sparse graphs.
\end{itemize}

\begin{figure*}[htbp]
    \centering
    % The architecture figure now takes up a significant portion of the page width
    \includegraphics[width=0.8\textwidth]{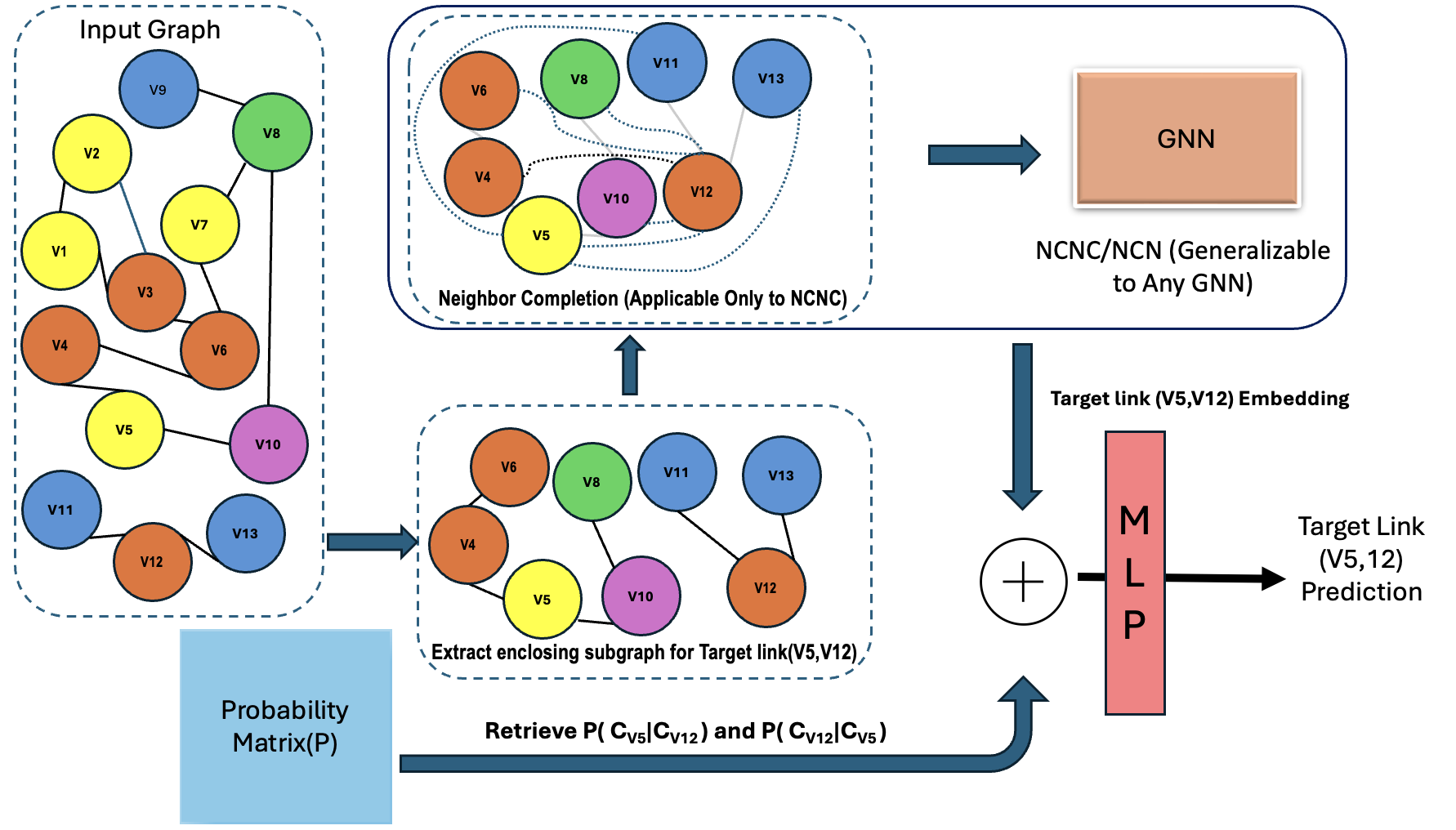}
    \caption{The CGLE architecture for the task of link prediction, exemplified by the target link \(V_5 \leftrightarrow V_{12}\). The underlying backbone model is NCN/NCNC, which incorporates subgraph extraction and, in the case of NCNC, a neighbor completion module. The broader CGLE framework is designed to be compatible with various GNN-based architectures for link prediction.}
    \label{fig:cgle_arch}
\end{figure*}

\begin{prop}
Extending Theorem 1, we propose a refined class-label-integrated heuristic that combines both structural information and class-label interactions:

\begin{equation}
    H_C(x, y) = H(x, y) + \beta \cdot \frac{\alpha_1 \operatorname{P}\left(c_y \mid c_x\right) + \alpha_2 \operatorname{P}\left(c_x \mid c_y\right)}{Z(x, y)}
\label{eq:class_heuristic}
\end{equation}

Here, \(H(x, y)\) represents structural information as a \(\gamma\)-decaying heuristic for nodes \(x\) and \(y\) with class labels \(c_x\) and \(c_y\). The conditional probabilities for inter-class linkage, \(P(c_y \mid c_x)\) and \(P(c_x \mid c_y)\), are detailed in Equation~\ref{eq:equation7}. These probabilities form the entries of the matrix \(P\), which is visualized for several datasets in Fig.~\ref{fig:heatmap_analysis_2x3}.

The learnable parameters \(\alpha_1\) and \(\alpha_2\) control the relative contributions of these probabilities and are optimized via an MLP.

The parameter \(\beta \in \mathbb{R}^+\) controls the contribution of class-label information, ensuring a balanced integration of structural and semantic factors.

The local normalization factor \(Z(x, y)\) is defined as:

\begin{equation}
    Z(x, y) = \sum_{v \in N(x) \cup N(y)} \sum_{i \in \{x, y\}} \left[ \omega_{1i} P(c_i \mid c_v) + \omega_{2i} P(c_v \mid c_i) \right]
\end{equation}

where \(\omega_{1i}, \omega_{2i} \in \mathbb{R}^+\) are additional learnable parameters optimized via an MLP. Here, \(N(x)\) and \(N(y)\) denotes the 1-hop neighbors of node \(x\) and \(y\).

In cases where only global class interaction information is required, the normalization factor can be simplified to:

\begin{equation}
\label{eq:eqation 4}
    Z(x, y) = 1.
\end{equation}
\end{prop}

\noindent\textbf{Conditional Probability Justification.}
\label{sec: justification}
The probability of a node in class \(c_x\) forming a link with a node in class \(c_y\) can be interpreted as a conditional probability:

\begin{equation}
    P(c_y \mid c_x) = \frac{P(c_x, c_y)}{P(c_x)}
    \label{equation:eq5}
\end{equation}

Here, the joint probability \(P(c_x, c_y)\) and the marginal probability \(P(c_x)\) are defined as:

\begin{align}
\label{eq:equation6}
    P(c_x, c_y) = \frac{\text{count}(c_x \rightarrow c_y)}{N} \quad \text{and} \\ 
    P(c_x) = \frac{\sum_{c \in C} \text{count}(c_x \rightarrow c)}{N}
\end{align}

where \( N \) denotes the total number of edges in the graph. Substituting these definitions into Equation \ref{equation:eq5}, we obtain:

\begin{equation}
\label{eq:equation7}
    P(c_y \mid c_x) = \frac{\text{count}(c_x \rightarrow c_y)}{\sum_{c \in C} \text{count}(c_x \rightarrow c)}
\end{equation}
In the above equations (\ref{eq:equation6} -- \ref{eq:equation7}), \( C \) denotes the set of all possible classes, and \( c \in C \) represents an individual class within this set.
This formulation emphasizes the conditional nature of the probability, ensuring that:
\begin{itemize}
    \item \(P(c_y \mid c_x)\) represents the likelihood of a node in class \(c_x\) linking to class \(c_y\).
    \item The denominator accounts for all outgoing edges from \(c_x\), naturally capturing asymmetric connectivity patterns.
\end{itemize}

Under the conditions stated in Theorem 1, if the structural term \(f(x, y, l)\) follows the same exponential decay conditions, the extended heuristic \(H_C(x, y)\) retains the exponential error bound properties when approximated from an \(h\)-hop enclosing subgraph. This ensures that the refined heuristic maintains theoretical soundness while enhancing predictive performance through class-label integration.

The redefined formulation addresses the limitations of traditional heuristics, which struggle in sparse or heterophilous graphs. By integrating class-label information, it enhances link prediction by capturing global class-based patterns.

In Fig.~\ref{fig:link_prediction_example}, nodes \(V_5\) and \(V_{12}\) are disconnected in the training graph, making conventional methods like Common Neighbors (CN) ineffective. The redefined formulation overcomes this by leveraging class-label dependencies, enabling models like CGLE to predict potential links despite missing structural cues. This improves disambiguation, reveals long-range dependencies, and mitigates data sparsity, enhancing predictive accuracy in complex graphs.

\section{CGLE: An Implementation of Theory}
\label{sec: implemtation}

Equation~\ref{eq:class_heuristic} provides the theoretical foundation for CGLE, modeling the probability of a link between two nodes as a weighted combination of structural similarity $H(x, y)$ and class-conditioned link prior $P(c_x \mid c_y)$ and  $P(c_y \mid c_x)$, controlled by coefficients $\alpha$ and $\beta$. While this formulation offers valuable conceptual intuition, it is not directly applied in the implementation.

Instead, the practical model replaces the explicit weighting with a trainable function. Specifically, we concatenate the node embeddings with the estimated class-conditioned probabilities, $P(c_y \mid c_x)$ and $P(c_x \mid c_y)$, and input this composite vector to a multi-layer perceptron (MLP), as expressed in Equation~\ref{eq: eqation 12}. This design eliminates the need to manually tune $\alpha$ and $\beta$, allowing the MLP to learn an optimal, potentially non-linear, fusion of structural and semantic features.

In this section, we describe how the theoretical motivations are operationalized in the CGLE framework. We detail the model architecture, the construction of the probability matrix $P$, and the steps involved in performing class-aware link prediction across both labeled and unlabeled graphs.

% \begin{figure}[t]
%     \centering
%     \includegraphics[width=0.7\linewidth]{Paper_figures/prob_matrix.png} 
%     \caption{The probability matrix \( P \) is computed with a time complexity of \( O(E) \), where \( E \) denotes the total number of training edges.}
%     \label{fig:figure2}
% \end{figure}

\subsection{Inter-class Link Probability Calculation}
We compute a probability matrix \( P \) of size \( |C| \times |C| \), where \( |C| \) represents the total number of unique classes. Each entry in \( P \) corresponds to the link formation probability between pairs of class groups, calculated using Equation~\ref{equation:eq5}.

The computed probability matrix \( P \) is then incorporated into the final multi-layer perceptron (MLP)~\cite{rumelhart1986learning} by concatenating with the NCNC~\cite{ncnc2024} link embedding, resulting in notable improvements in link prediction accuracy.

\subsection{Graph Message Passing Framework}
We consider an undirected graph $G = (V, E, A, X)$, where $V = \{1, 2, \dots, n\}$ represents a set of $n$ nodes, and $E \subseteq V \times V$ is the set of edges between them. The matrix $X \in \mathbb{R}^{n \times F}$ is the node feature matrix, where each row $X_v$ contains the feature vector for node $v$. The adjacency matrix $A \in \mathbb{R}^{n \times n}$ is symmetric, with $A_{uv} = 1$ if nodes $u$ and $v$ are connected, i.e., $(u, v) \in E$, and $0$ otherwise.

The degree of node $u$ is defined as $d(u, A) = \sum_{v=1}^{n} A_{uv}$, which counts the number of edges incident to node $u$. The set of neighbors of a node $u$, denoted by $N(u, A)$, consists of all nodes $v$ such that $A_{uv} > 0$. For simplicity, when the adjacency matrix $A$ is fixed, we refer to this neighborhood as $N(u)$.

For two nodes $x$ and $y$, their common neighbors are represented by $N(x) \cap N(y)$, indicating the set of nodes that are connected to both $x$ and $y$. The difference in their neighborhoods is given by $N(y, A) - N(x, A)$, which represents nodes connected to $y$ but not to $x$, and similarly, $N(x, A) - N(y, A)$ represents nodes connected to $i$ but not to $j$.

\textbf{Message Passing Neural Network (MPNN).} A widely used GNN framework, MPNN~\cite{gilmer2017neural}, consists of multiple message-passing layers that propagate information between nodes. At the $k$-th layer, the node representation for node $v$ is updated as follows:
\begin{equation}
\small
h_v^{(k)} = U^{(k)}\left(h_v^{(k-1)}, \text{AGG}\left(\{M^{(k)}(h_v^{(k-1)}, h_u^{(k-1)}) \mid u \in N(v)\}\right)\right),
\label{eq:mpnn}
\end{equation}
where $h_v^{(k)}$ represents the embedding of node $v$ at layer $k$, $U^{(k)}$ and $M^{(k)}$ are learnable functions (often implemented as multi-layer perceptrons), and $\text{AGG}$ is an aggregation function, such as summing or taking the maximum over the messages from the neighboring nodes. Initially, the node representations are set as $h_v^{(0)} = X_v$, i.e., the node feature vectors. After $K$ layers of message passing, the final node representations are denoted as $\text{MPNN}(v, A, X) = h_v^{(K)}$.

% \begin{figure*}[htbp]
%     \centering
%     \includegraphics[width=0.7\textwidth]{Paper_figures/architechture.png} 
%     \caption{CGLE architecture illustrating the prediction of a target link (e.g., \( V_5 \leftrightarrow V_{12} \)). The process involves subgraph extraction, probability matrix integration, GNN-based embedding generation, and MLP-based prediction.}
%     \label{fig:cgle_arch}
% \end{figure*}

\subsection{Model Architecture}

\subsubsection{Common Neighbor Completion (CNC)}
The implementation extends the \textit{Neural Common Neighbor with or without Completion (NCN or NCNC)} model~\cite{ncnc2024}, which employs the \textit{Common Neighbor Completion (CNC)} technique to address graph incompleteness. Rather than attempting to reconstruct the entire graph, CNC focuses on selectively completing common neighbor links, ensuring improved efficiency, particularly for large-scale graph datasets.

For a specific node pair \( (x, y) \), we define the probability \( P_{uxy} \) that a node \( u \) serves as a common neighbor for the pair as follows:

\begin{equation}
P_{uxy} =
\begin{cases}
1 & \text{if } u \in N(x, A) \cap N(y, A) \\
\hat{A}_{xu} & \text{if } u \in N(y, A) - N(x, A) \\
\hat{A}_{yu} & \text{if } u \in N(x, A) - N(y, A) \\
0 & \text{otherwise}
\end{cases}
\label{eq:probability_common_neighbor}
\end{equation}

In this context, \( \hat{A}_{xu} \) denotes the predicted probability of the existence of the link \( (x, u) \). The model assumes that \( u \) qualifies as a common neighbor of \( (x, y) \) if both edges \( (x, u) \) and \( (y, u) \) are present. When either edge is unobserved, the model utilizes NCN to predict the likelihood of \( u \) being a common neighbor. If both edges are absent, the probability is assigned a value of 0.

Once the common neighbor links are completed, we reapply the NCN model to the modified graph. The final formulation of the \textit{Neural Common Neighbor with Completion (NCNC)} model is expressed as follows:
\begin{align}
\text{NCN}(x, y, A, X) &= 
\text{MPNN}(x, A, X) \odot \text{MPNN}(y, A, X) \,\| \nonumber \\
&\quad \sum_{u \in N(x) \cap N(y)} P_{uxy} \cdot \text{MPNN}(u, A, X) \\
\text{NCNC}(x, y, A, X) &= 
\text{MPNN}(x, A, X) \odot \text{MPNN}(y, A, X) \,\| \nonumber \\
&\quad \sum_{u \in N(x) \cup N(y)} P_{uxy} \cdot \text{MPNN}(u, A, X)
\label{eq:equation 10}
\end{align}

In these equations, \( \odot \) represents element-wise multiplication, and \( \parallel \) indicates concatenation. The summation aggregates information from the predicted common neighbors, weighted by their respective probabilities.

\subsubsection{\textbf{CGLE}} 
Our proposed method, \textit{CGLE} (Fig. \ref{fig:cgle_arch}), is designed to enhance link prediction by integrating class-label information with structural features. Although our implementation builds upon NCN and NCNC, the framework is flexible and can seamlessly incorporate any GNN model. 

In CGLE, we leverage class-conditioned probabilities in conjunction with node embeddings to improve predictive accuracy. For each node pair \( (x, y) \), the final link prediction score is computed as follows: 
\begin{equation}
\label{eq: eqation 12}
y_{xy} = \text{MLP} \left( f(x, y, A, X) \parallel P(c_y \mid c_x) \parallel P(c_x \mid c_y) \right),
\end{equation}
where, \(f(x, y, A, X)\) represents a flexible function that can capture the output of NCN, NCNC, or any other GNN model, ensuring adaptability across diverse graph learning architectures. When \(f\) corresponds to NCNC, it aligns with the NCNC model’s prediction for nodes \(x\) and \(y\) (Eq. \ref{eq:equation 10}). 

The terms \(P(c_y \mid c_x)\) and \(P(c_x \mid c_y)\) (Eq. \ref{equation:eq5}) denote the conditional probabilities of link formation between the corresponding class labels. By combining these class-conditioned probabilities with structural node embeddings, CGLE effectively integrates both structural and semantic information to enhance link prediction.

\subsection{{Generating Pseudo Class Labels for Unlabeled Graphs}}
\label{sec: psudo_class}
In the absence of true class labels, we propose a method that leverages graph structure and node features to generate pseudo-labels. This ensures meaningful node groupings, enhancing class-based link prediction.

\subsubsection{Step 1: 1-Hop Neighborhood Aggregation} 
To construct node embeddings, we aggregate features from each node’s 1-hop neighbors, capturing local graph structure: 
\begin{equation}
H^1 = (Adj_{\text{train}} + I)X
\label{eq:hop1-agg}
\end{equation}
Here, \( \textit{Adj}_{\textit{Train}} \) is the adjacency matrix of the training subgraph, \( I \) is the identity matrix, and \( X \) is the node feature matrix. Each row \( H_v \) represents the combined features of node \( v \), embedding both node attributes and neighborhood structure.

\subsubsection{Step 2: k-means Clustering} 
We then apply k-means clustering~\cite{lloyd1982least} to partition nodes into \( k \) clusters. This groups nodes with similar features and structure. The algorithm iteratively minimizes intra-cluster variance, producing stable pseudo-labels for link prediction tasks.

\subsubsection{Step 3: Optimal Cluster Selection (\( k \))} 
To determine the optimal \( k \), we use the \textit{elbow method}, which identifies the point where the rate of variance reduction slows. This optimal \( k \) value improves clustering performance and enhances class-aware link prediction.

\begin{figure}[t]
    \centering
    \includegraphics[width=0.9\linewidth]{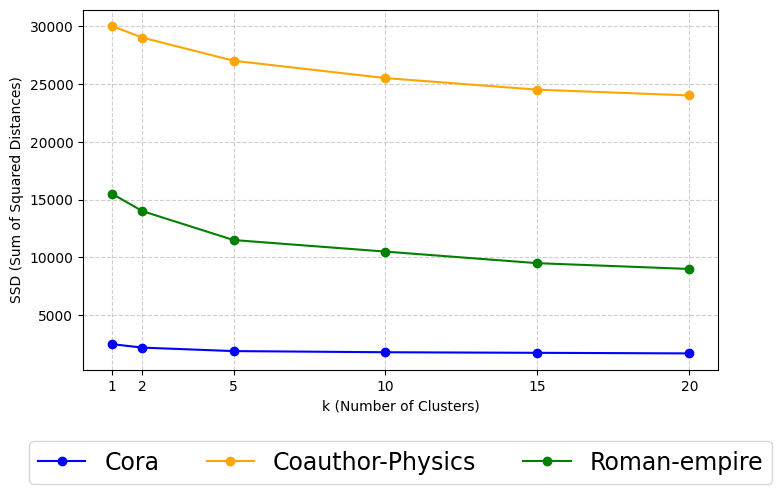}
    \caption{Elbow plots for determining the optimal number of clusters (\(k\)). The plots illustrate the Sum of Squared Distances (SSD) for varying \(k\) on the \textcolor{blue}{Cora}, \textcolor{orange}{Coauthor-Physics}, and \textcolor{deepgreen}{Roman-empire} datasets.}
    \label{fig:elbow}
\end{figure}

\subsection{The CGLE Algorithm}

This section outlines the CGLE algorithm, an extension of the NCNC model that leverages class-based probabilities for enhanced link prediction. The algorithm's procedure is detailed in Algorithm \ref{alg:CGLE}. When class labels are not available, an optimal number of clusters (\(k\)) for pseudo-labeling is determined using the elbow method, as illustrated in Fig.~\ref{fig:elbow}.

\begin{algorithm}[htbp]
\small
\caption{CGLE: Class-label Graph Link Estimator}
\label{alg:CGLE}

\SetKwInOut{Input}{Input}
\SetKwInOut{Output}{Output}

\Input{
    Graph $G=(V, E)$; \\
    Node class labels $C = \{c_1, c_2, \dots, c_n\}$; \\
    A backbone GNN function $f(x, y, A, X)$;
}
\Output{Final link probability $y_{xy}$ for any node pair $(x, y)$}

\BlankLine

\SetKwComment{Comment}{$\triangleright$\ }{}
\Comment{Phase 1: Learn class-level link probabilities (one-time pre-computation)}
Group all nodes $V$ by their class labels in $C$\;
Compute the class-conditioned probability matrix $P$, where entries $P(c_y \mid c_x)$ are the conditional probabilities for inter-class linkage, where nodes $x$ and $y$ belong to class labels $c_x$ and $c_y$\;

\BlankLine

\Comment{Phase 2: Making the final prediction for a node pair $(x, y)$}
\SetKwFunction{FMain}{PredictLink}
\SetKwProg{Fn}{Function}{}{}

\Fn{\FMain{$x, y$}}{
    \Comment{Get the structural embedding from the backbone GNN}
    $H(x,y) \leftarrow f(x, y, A, X)$\;
    
    \Comment{Look up the pre-computed class-level priors}
    $P(c_y \mid c_x) \leftarrow P[c_y][c_x]$\;
    $P(c_x \mid c_y) \leftarrow P[c_x][c_y]$\;
    
    \Comment{Concatenate structural and class-level features}
    $e_{\text{combined}} \leftarrow \text{concat}(H(x,y), P(c_x \mid c_y), P(c_y \mid c_x))$\;
    
    \Comment{Make the final prediction using an MLP}
    $y_{xy} \leftarrow \text{MLP}(e_{\text{combined}})$\;
    
    \KwRet $y_{xy}$\;
}
\end{algorithm}

\subsubsection{Complexity Analysis}
The computational complexity of CGLE's preprocessing phase depends on whether class labels are provided.

\begin{figure}[t]
    \centering
    \includegraphics[width=0.95\linewidth]{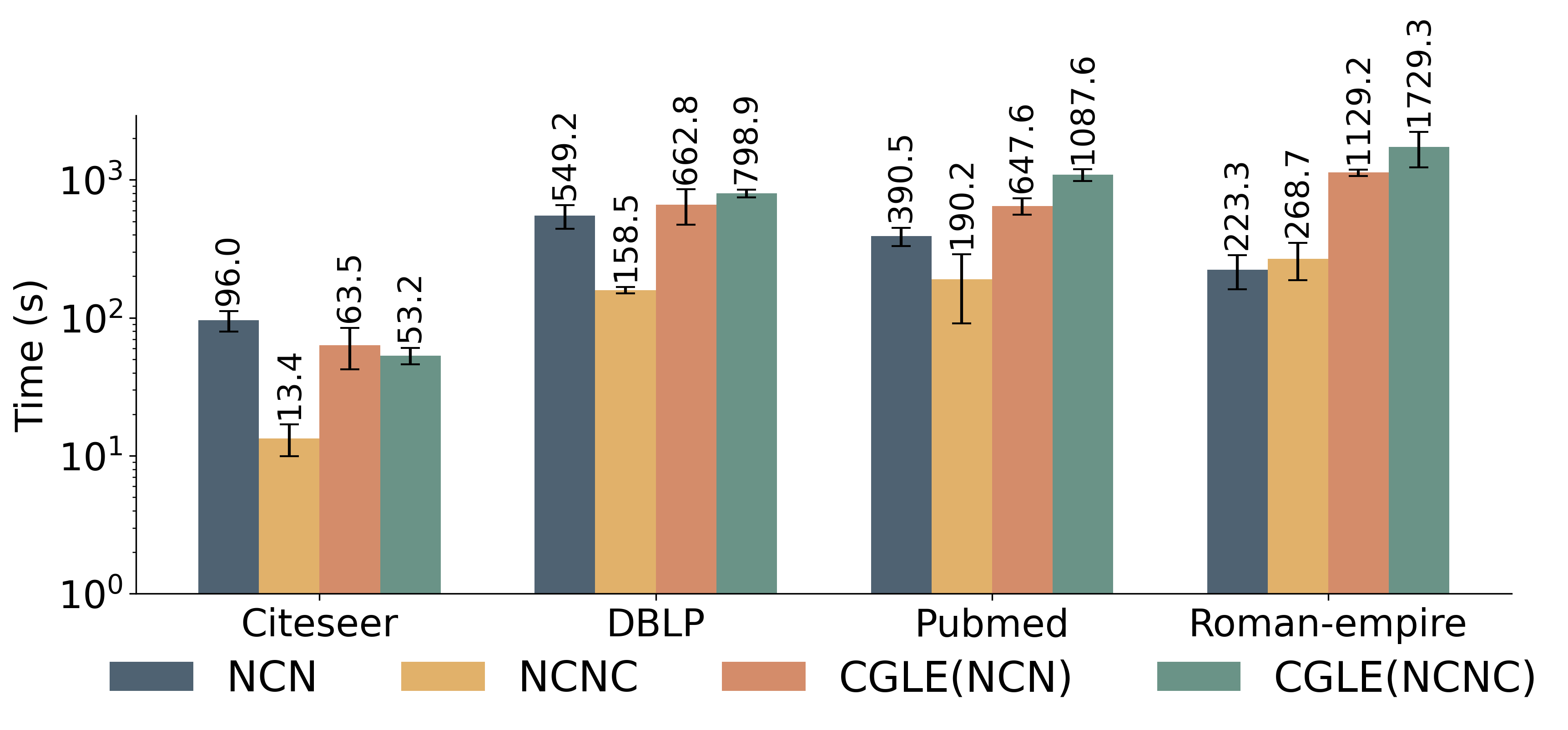}
    \caption{Execution time (in seconds) for the \textcolor{cglenchcolor}{CGLE(NCN)} and \textcolor{cglencnccolor}{CGLE(NCNC)} models compared to the \textcolor{ncncolor}{NCN} and \textcolor{ncnccolor}{NCNC} baselines. For brevity, this plot shows runtime on four selected datasets: Citeseer, DBLP, Pubmed and Roman-empire.}
    \label{fig:exection_time}
\end{figure}

\textbf{With available class labels}, the primary computational cost is the construction of the probability matrix \(P\). This process involves a single pass over the training edges, leading to a time complexity of \(\mathcal{O}(E)\), where \(E\) is the total number of edges. The execution times for several datasets are presented in Fig.~\ref{fig:exection_time}.

\textbf{Without class labels}, pseudo-labels must be generated through additional steps. First, aggregating neighborhood features to form matrix \(H^1\) (as per Eq. \ref{eq:hop1-agg}) has a complexity of \(\mathcal{O}(E \cdot F)\). Second, applying k-means clustering to these features takes \(\mathcal{O}(k \cdot V \cdot t)\) time, where \(k\) is the cluster count and \(t\) is the number of iterations. Consequently, the total preprocessing complexity in this scenario becomes \(\mathcal{O}(E + E \cdot F + k \cdot V \cdot t)\). Note that using the elbow method to find an optimal \(k\) would require running the clustering step multiple times, further adding to this computational cost.

\section{Experiments}
\label{sec: experimentation}
Our implementation choices and hyperparameter configurations are guided by the methodology proposed by Li et al.~\cite{li2023evaluatinggraphneuralnetworks}. For all experiments, we employed a single NVIDIA A100 GPU equipped with 80 GB of HBM2e memory.
\subsection{Datasets}
We evaluate our method on several popular graph datasets, categorized into two types: \textit{homophilous} and \textit{heterophilous} graphs, as shown in Table~\ref{tab:table1_single_col}.

For the homophilous graphs, including Cora~\cite{Yang2016RevisitingSL}, Citeseer~\cite{Yang2016RevisitingSL}, Pubmed~\cite{Yang2016RevisitingSL}, Facebook~\cite{Yang2020PANESA}, Coauthor-Physics~\cite{shchur2019pitfallsgraphneuralnetwork}, DBLP~\cite{bojchevski2018deepgaussianembeddinggraphs}, and FB Page-Page~\cite{rozemberczki2021multiscaleattributednodeembedding}.

For the heterophilous graphs, including Roman-empire, Amazon-ratings, Questions~\cite{Platonov2023ACL}, Actor~\cite{Pei2020Geom-GCN:} and Chameleon~\cite{rozemberczki2021multiscaleattributednodeembedding}. These datasets are selected to evaluate the robustness of our method under challenging structural conditions where connected nodes often belong to different classes.
\begin{table}[tbp]
\caption{Dataset statistics for homophilous and heterophilous graphs. The data split is 85\% training, 5\% validation, and 10\% testing.}
\centering
% Use a smaller font and reduce column padding to save space
\footnotesize
\setlength{\tabcolsep}{3pt}
\begin{tabular}{l l r r r r}
\toprule
\textbf{Type} & \textbf{Dataset} & \textbf{Nodes} & \textbf{Edges} & \textbf{Feats.} & \textbf{Classes} \\
\midrule
\multirow{7}{*}{\textit{Homophilous}} 
 & Cora           & 2,708  & 10,556  & 1,433 & 7   \\
 & Citeseer       & 3,327  & 9,104   & 3,703 & 6   \\
 & Pubmed         & 19,717 & 88,648  & 500   & 3   \\
 & FB Page-Page   & 22,470 & 171,002 & 31    & 4   \\
 & Coauthor-Physics & 34,493 & 495,924 & 8,415 & 5   \\
 & Facebook       & 4,039  & 88,234  & 1,283 & 193 \\
 & DBLP           & 17,716 & 105,734 & 1,639 & 4   \\
\cmidrule(lr){1-6}
\multirow{6}{*}{\textit{Heterophilous}} 
 & Roman-empire   & 22,662 & 32,927  & 300   & 18  \\
 & Amazon-ratings & 24,492 & 93,050  & 300   & 5   \\
 & Questions      & 48,921 & 153,540 & 301   & 2   \\
 & Chameleon      & 2,277  & 36,101  & 2,325 & 5   \\
 & Actor          & 7,600  & 33,544  & 931   & 5   \\
\bottomrule
\end{tabular}
\label{tab:table1_single_col}
\end{table}
\begin{table*}[t]
\caption{Link prediction performance on benchmark homophilous datasets. The top three results are highlighted: \textcolor{blue}{1st}, \textcolor{deepgreen}{2nd}, and \textcolor{orange}{3rd} highest scores in each column. For the optimal $ k $ value, see Fig. \ref{fig:k_means}.}
\centering
\begin{tabular}{l@{\hskip 12pt}c@{\hskip 12pt}c@{\hskip 12pt}c@{\hskip 12pt}c@{\hskip 12pt}c@{\hskip 12pt}c@{\hskip 12pt}c}
\toprule
\multirow{2}{*}{\textbf{Method}} & \textbf{Cora} & \textbf{Citeseer} & \textbf{Pubmed} & \textbf{FB Page-Page} & \textbf{Facebook} & \textbf{Coauthor-Physics} & \textbf{DBLP} \\
 & HR@100 & HR@100 & HR@100 & MRR & HR@100 & MRR & HR@10 \\
\midrule
\textbf{CN} & 33.92 & 29.79 & 23.13 & 17.85 & 84.38 & 18.57 & 32.8 \\
\textbf{AA} & 39.85 & 35.19 & 27.38 & 22.6 & 88.14 & 22.31 & 21.13 \\
\textbf{RA} & 41.07 & 33.56 & 27.03 & 20.54 & 92.58 & 21.46 & 22.47 \\
\cmidrule(lr){1-8}
\textbf{GCN} & 66.79$\pm$1.65 & 67.08$\pm$2.94 & 53.02$\pm$1.39 & 11.26$\pm$1.6 & 92.85$\pm$0.61 & 14.68$\pm$3.40 & 33.30$\pm$4.74 \\
\textbf{SAGE} & 55.02$\pm$4.03 & 57.01$\pm$3.57 & 44.29$\pm$1.44 & 10.44$\pm$2.48 & 68.50$\pm$8.6 & 13.07$\pm$1.02 & 31.06$\pm$5.98 \\
\textbf{GAE} & 89.01$\pm$1.32 & 91.78$\pm$0.94 & 78.81$\pm$1.64 & 12.93$\pm$0.66 & 92.68$\pm$2.58 & 15.83$\pm$1.67 & 41.38$\pm$3.72 \\
\cmidrule(lr){1-8}
\textbf{Neo-GNN} & 80.42$\pm$1.34 & 84.67$\pm$1.42 & 73.93$\pm$1.19 & 12.43$\pm$0.22 & 91.24$\pm$0.77 & 20.94$\pm$3.94 & 50.05$\pm$3.40 \\
\textbf{BUDDY} & 88.00$\pm$0.44 & 92.93$\pm$0.27 & 74.10$\pm$0.78 & \textcolor{deepgreen}{16.94$\pm$1.37} & 87.56$\pm$1.43 & 14.26$\pm$1.82 & 31.74$\pm$6.09 \\
\textbf{NCN} & 89.05$\pm$0.96 & 91.56$\pm$1.43 & 79.05$\pm$1.16 & 9.16$\pm$1.96 & 93.67$\pm$0.82 & \textcolor{blue}{29.05$\pm$3.48} & {51.26$\pm$3.26} \\
\textbf{NCNC} & 89.65$\pm$1.36 & 93.47$\pm$0.95 & 81.29$\pm$0.85 & 14.03$\pm$7.88 & 92.78$\pm$2.00 & 20.99$\pm$5.09 & 42.82$\pm$4.12 \\
\cmidrule(lr){1-8}
\textbf{NCN $\parallel$ True Class Label} & \textcolor{deepgreen}{95.71$\pm$1.10} & \textcolor{deepgreen}{96.96$\pm$0.37} & \textcolor{deepgreen}{90.81$\pm$1.13} & 11.27$\pm$4.62 & {93.69$\pm$0.62} & \textcolor{orange}{27.04$\pm$3.93} & \textcolor{deepgreen}{51.75$\pm$2.55} \\
\textbf{CGLE(NCN)(True Class Label)} & \textcolor{blue}{95.77$\pm$0.62} & \textcolor{blue}{97.27$\pm$0.74} & \textcolor{orange}{90.49$\pm$0.54} & 12.06$\pm$5.57 & \textcolor{orange}{93.75$\pm$0.79} & {26.97$\pm$4.32} & \textcolor{orange}{51.33$\pm$2.00} \\
\textbf{NCNC $\parallel$ True Class Label} & 88.63$\pm$1.72 & 92.46$\pm$1.05 & 82.02$\pm$1.51 & 12.72$\pm$8.41 & 92.95$\pm$0.62 & 21.48$\pm$6.47 & 42.54$\pm$4.28 \\
\textbf{CGLE(NCNC)(True Class Label)} & 91.41$\pm$1.36 & 92.31$\pm$0.14 & 82.06$\pm$0.13 & \textcolor{blue}{23.84$\pm$6.15} & \textcolor{deepgreen}{93.92$\pm$0.56} & 21.24$\pm$3.06 & 49.00$\pm$3.10 \\
\textbf{CGLE(NCN)-k-means (Best k)}& {94.27$\pm$0.94} & {95.89$\pm$1.84} & {90.44$\pm$0.83} & {7.84$\pm$1.28} & \textcolor{blue}{93.99$\pm$0.59} & \textcolor{deepgreen}{27.29$\pm$3.47} & \textcolor{blue}{52.86$\pm$1.48} \\
\textbf{CGLE(NCNC)-k-means (Best k)}& \textcolor{orange}{94.80$\pm$0.96} & \textcolor{orange}{96.90$\pm$1.12} & \textcolor{blue}{91.65$\pm$0.60} & \textcolor{orange}{16.32$\pm$5.70} & 93.61$\pm$0.90 & 24.94$\pm$4.42 & 48.88$\pm$3.21 \\
\bottomrule
\end{tabular}
\label{tab:table 2}
\end{table*}

\begin{table*}[t]
\caption{Link prediction performance on benchmark heterophilous datasets. The top three results are highlighted: \textcolor{blue}{1st}, \textcolor{deepgreen}{2nd}, and \textcolor{orange}{3rd} highest scores in each column.}
\centering
\begin{tabular}{lc@{\hskip 12pt}c@{\hskip 12pt}c@{\hskip 12pt}c@{\hskip 12pt}c}
\toprule
\multirow{2}{*}{\textbf{Method}} & \textbf{Roman-empire} & \textbf{Amazon-ratings} & \textbf{Questions} & \textbf{Chameleon} & \textbf{Actor} \\
\cmidrule{2-6}
 & MRR & MRR & HR@100 & MRR & HR@100 \\
\midrule
\textbf{NCN} & \textcolor{blue}{54.29 $\pm$ 0.86} & 55.90 $\pm$ 7.51 & 62.25 $\pm$ 1.75 & 76.79 $\pm$ 1.33 & \textcolor{orange} {53.18 $\pm$ 1.65} \\
\textbf{NCNC} & 28.23 $\pm$ 12.51 & \textcolor{deepgreen} {72.63 $\pm$ 6.69} & 62.93 $\pm$ 1.73 & 74.75 $\pm$ 8.37 & 50.77 $\pm$ 3.07 \\
\cmidrule(lr){1-6}
\textbf{NCN $\parallel$ Class Label} & 52.32 $\pm$ 1.96 & 59.88 $\pm$ 8.72 & \textcolor{deepgreen} {63.89 $\pm$ 1.40} & 77.09 $\pm$ 2.92 & 51.01 $\pm$ 2.35 \\
\textbf{NCNC $\parallel$ Class Label} & 32.35 $\pm$ 11.88 & 67.56 $\pm$ 3.17 & \textcolor{deepgreen} {63.89 $\pm$ 1.40} & 73.68 $\pm$ 7.78 & 51.48 $\pm$ 1.19 \\
\cmidrule(lr){1-6}
\textbf{CGLE(NCN)(True Class Label)} & \textcolor{deepgreen} {54.01 $\pm$ 0.71} & 64.68 $\pm$ 8.25 & 63.02 $\pm$ 1.55 & \textcolor{blue}{81.15 $\pm$ 3.09} & \textcolor{deepgreen} {53.37 $\pm$ 1.71} \\
\textbf{CGLE(NCNC)(True Class Label)} & 52.23 $\pm$ 2.31 & \textcolor{orange} {70.62 $\pm$ 5.96} & \textcolor{orange} {63.44 $\pm$ 1.57} & \textcolor{deepgreen} {77.88 $\pm$ 8.29} & 51.07 $\pm$ 4.31 \\
\textbf{CGLE(NCN)-k-means (Best k)} & 53.19 $\pm$ 1.44 & {64.03 $\pm$ 6.87} & {61.33 $\pm$ 2.98} & 77.32 $\pm$ 4.19 & \textcolor{blue}{54.82 $\pm$ 1.57} \\
\textbf{CGLE(NCNC)-k-means (Best k)} & \textcolor{orange} {53.82 $\pm$ 2.57} & \textcolor{blue}{73.67 $\pm$ 5.11} & \textcolor{blue}{63.95 $\pm$ 2.82} & \textcolor{orange} {77.87 $\pm$ 5.45} & 51.42 $\pm$ 3.87 \\
\bottomrule
\end{tabular}
\label{tab:table 3}
\end{table*}

We compute the link probabilities between nodes, using only the training edges, and concatenate these probabilities with the NCNC embeddings to enhance the link prediction task. 
In our experiments, as presented in Tables \ref{tab:table 2} and \ref{tab:table 3} we evaluate three distinct setups: Table \ref{tab:table 2} focuses on homophilic datasets, while Table \ref{tab:table 3} presents results for heterophilic datasets.

\begin{figure*}[t!]
\centering
% The single \includegraphics command replaces all the previous \subfloat commands.
% Make sure the path to the image is correct for your project structure.
\includegraphics[width=\linewidth]{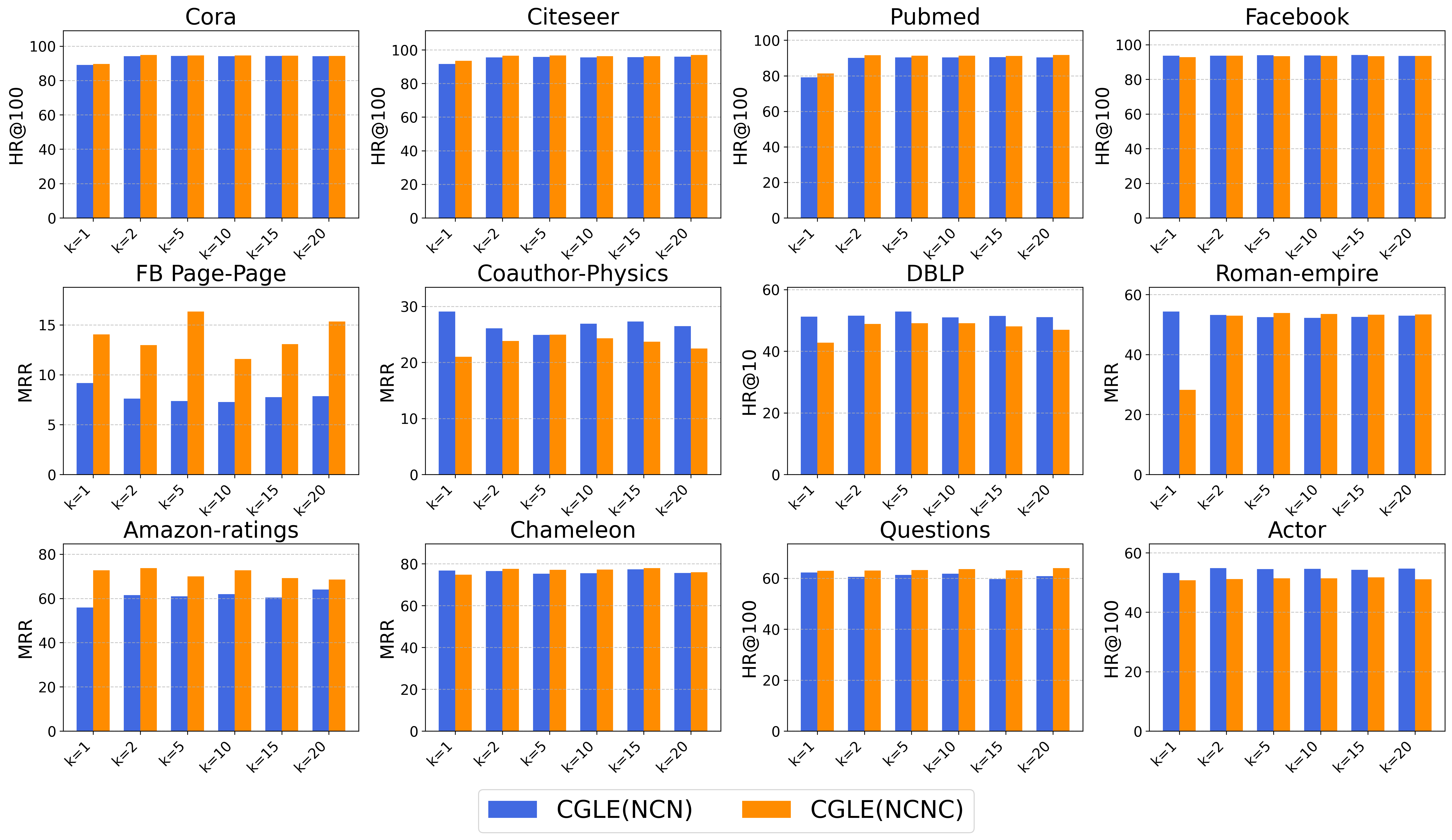}

\caption{Link prediction performance of CGLE, using NCN and NCNC backbones, across 12 datasets for different numbers of k-means clusters ($k \in \{1, 2, 5, 10, 15\}$). The $k=1$ case serves as a baseline, equivalent to running the backbone models without class labels. The first seven datasets are homophilous, and the remaining five are heterophilous. Each subplot shows a specific performance metric (HR@100, MRR, or HR@10) for one dataset.}
\label{fig:k_means}

\end{figure*}

\begin{itemize}
    \item \textbf{True Class Labels (CGLE)}: In this setting, the true class labels available in the datasets are incorporated during both training and evaluation, enriching the model with class-awareness. This setup corresponds to CGLE(NCN)(True Class Label) and CGLE(NCNC)(True Class Label), where NCN and NCNC are the respective backbone GNN models.
    \item \textbf{Generated Class Labels (CGLE(NCNC/NCN)-k-means)}: When true class labels are unavailable, pseudo labels are generated using k-means clustering. This setup is denoted as CGLE(NCNC/NCN)-k-means (Best k) based on the backbone model. Detailed results and analysis can be found at Fig.~\ref{fig:k_means}. For a comprehensive explanation of the proposed method, see section \ref{sec: psudo_class}.
    \item \textbf{NCN $\parallel$ True Class Labels} and \textbf{NCNC $\parallel$ True Class Labels}: In these configurations, a one-hot encoded vector of true class labels is concatenated (denoted by $\parallel$) with the raw node embeddings. This augmented feature representation aims to boost predictive performance by integrating class information directly into the embeddings.
\end{itemize}

\begin{table*}[htbp]
\centering
\caption{Ablation study of the CGLE framework on homophilous datasets. This table compares the performance of CGLE using NCN and NCNC as backbones against the prior best results from Table~\ref{tab:table 2}.}
\label{table:tab 4}
\begin{tabular}{@{}llccccccc@{}}
\toprule
\textbf{Variant} & \textbf{Method} & \begin{tabular}[c]{@{}c@{}}\textbf{Cora}\\\textit{HR@100}\end{tabular} & \begin{tabular}[c]{@{}c@{}}\textbf{Citeseer}\\\textit{HR@100}\end{tabular} & \begin{tabular}[c]{@{}c@{}}\textbf{Pubmed}\\\textit{HR@100}\end{tabular} & \begin{tabular}[c]{@{}c@{}}\textbf{FB Page-Page}\\\textit{MRR}\end{tabular} & \begin{tabular}[c]{@{}c@{}}\textbf{Facebook}\\\textit{HR@100}\end{tabular} & \begin{tabular}[c]{@{}c@{}}\textbf{Coauthor-Physics}\\\textit{MRR}\end{tabular} & \begin{tabular}[c]{@{}c@{}}\textbf{DBLP}\\\textit{HR@10}\end{tabular} \\ \midrule
& Prior Best Result & 95.77$\pm$0.62 & 97.27$\pm$0.74 & \textbf{91.65$\pm$0.60} & \textbf{23.84$\pm$6.15} & 93.99$\pm$0.59 & \textbf{29.05$\pm$3.48} & \textbf{52.86$\pm$1.48} \\ \cmidrule(lr){2-9}
NCN & CGLE(Mono-Label) & 96.28 $\pm$ 0.41 & 97.16 $\pm$ 0.96 & 90.54 $\pm$ 0.67 & 11.01 $\pm$ 5.26 & \textbf{94.08 $\pm$ 0.36} & 26.90 $\pm$ 3.78 & 51.74 $\pm$ 1.77 \\ & CGLE(Louvain) & \textbf{96.32 $\pm$ 0.21} & 93.45 $\pm$ 6.69 & 80.62 $\pm$ 16.05 & 4.84 $\pm$ 1.49 & 91.26 $\pm$ 0.38 & 21.47 $\pm$ 2.12 & 52.15 $\pm$ 1.41 \\ & CGLE(Spectral) (best K) & 96.28 $\pm$ 0.63 & \textbf{97.76 $\pm$ 0.48} & 90.78 $\pm$ 0.72 & 11.70 $\pm$ 5.66 & 93.93 $\pm$ 0.53 & 28.53 $\pm$ 4.23 & 52.34 $\pm$ 2.75 \\ 
% & NCN & 89.05 $\pm$ 0.96 & 91.56 $\pm$ 1.43 & 79.05 $\pm$ 1.16 & 9.16 $\pm$ 1.96 & 93.67 $\pm$ 0.82 & 29.05 $\pm$ 3.48 & 51.26 $\pm$ 3.26 \\ 
\cmidrule(lr){1-9}
NCNC & CGLE (Mono-Label) & 95.41 $\pm$ 0.93 & 96.83 $\pm$ 0.50 & 91.40 $\pm$ 0.53 & 14.54 $\pm$ 5.46 & 93.40 $\pm$ 1.46 & 23.39 $\pm$ 4.90 & 48.95 $\pm$ 2.87 \\ & CGLE(Louvain) & 96.09 $\pm$ 1.01 & 96.77 $\pm$ 1.00 & 87.60 $\pm$ 0.59 & 4.69 $\pm$ 1.74 & 90.52 $\pm$ 0.56 & 20.17 $\pm$ 3.54 & 51.15 $\pm$ 1.63 \\ & CGLE(Spectral) (best K) & 96.00 $\pm$ 0.50 & 96.39 $\pm$ 1.21 & 87.78 $\pm$ 0.45 & 5.84 $\pm$ 2.54 & 90.98 $\pm$ 0.35 & 22.99 $\pm$ 4.85 & 52.25 $\pm$ 1.31 \\ 
% & NCNC & 89.65 $\pm$ 1.36 & 93.47 $\pm$ 0.95 & 81.29 $\pm$ 0.85 & 14.03 $\pm$ 7.88 & 92.78 $\pm$ 2.00 & 20.99 $\pm$ 5.09 & 42.82 $\pm$ 4.12 \\
\bottomrule
\end{tabular}
\end{table*}

\begin{table*}[htbp]
\centering
\caption{Ablation study of the CGLE framework on heterophilous datasets. This table compares the performance of CGLE using NCN and NCNC as backbones against the prior best results from Table~\ref{tab:table 3}.}
\label{table:tab 5}
\begin{tabular}{@{}llccccc@{}}
\toprule
\textbf{Variant} & \textbf{Method} & \begin{tabular}[c]{@{}c@{}}\textbf{Roman-empire}\\\textit{MRR}\end{tabular} & \begin{tabular}[c]{@{}c@{}}\textbf{Amazon-ratings}\\\textit{MRR}\end{tabular} & \begin{tabular}[c]{@{}c@{}}\textbf{Questions}\\\textit{HR@100}\end{tabular} & \begin{tabular}[c]{@{}c@{}}\textbf{Chameleon}\\\textit{MRR}\end{tabular} & \begin{tabular}[c]{@{}c@{}}\textbf{Actor}\\\textit{HR@100}\end{tabular} \\ \midrule
& Prior Best Result & 54.29 $\pm$ 0.86 & \textbf{73.67 $\pm$ 5.11} & \textbf{63.95 $\pm$ 2.82} & \textbf{81.15 $\pm$ 3.09} & \textbf{54.82 $\pm$ 1.57} \\ \cmidrule(lr){2-7}
NCN & CGLE (Mono-Label) & 54.01 $\pm$ 1.03 & 62.46 $\pm$ 6.39 & 61.96 $\pm$ 2.13 & 77.53 $\pm$ 2.86 & 53.04 $\pm$ 2.14 \\
& CGLE(Louvain) & 46.95 $\pm$ 0.80 & 43.91 $\pm$ 6.53 & 49.26 $\pm$ 6.02 & 57.86 $\pm$ 8.55 & 52.02 $\pm$ 0.65 \\
& CGLE(Spectral) (best K) & \textbf{54.57 $\pm$ 1.25} & 67.84 $\pm$ 7.11 & 62.89 $\pm$ 1.60 & 52.83 $\pm$ 3.57 & 50.93 $\pm$ 0.82 
% \\ & NCN & 54.29 $\pm$ 0.86 & 55.90 $\pm$ 7.51 & 62.25 $\pm$ 1.75 & 76.79 $\pm$ 1.33 & 53.18 $\pm$ 1.65 
\\ \cmidrule(lr){1-7}
NCNC & CGLE (Mono-Label) & 53.22 $\pm$ 2.19 & 70.38 $\pm$ 6.26 & 63.59 $\pm$ 1.66 & 74.94 $\pm$ 7.75 & 51.77 $\pm$ 2.27 \\
& CGLE(Louvain) & 48.97 $\pm$ 3.37 & 52.84 $\pm$ 7.24 & 51.41 $\pm$ 3.07 & 44.01 $\pm$ 10.37 & 52.00 $\pm$ 0.59 \\
& CGLE(Spectral) (best K) & 49.67 $\pm$ 4.24 & 52.49 $\pm$ 13.35 & 51.19 $\pm$ 3.65 & 47.70 $\pm$ 10.04 & 50.91 $\pm$ 1.77 \\
% & NCNC & 28.23 $\pm$ 12.51 & 72.63 $\pm$ 6.69 & 62.93 $\pm$ 1.73 & 74.75 $\pm$ 8.37 & 50.77 $\pm$ 3.07 \\
\bottomrule
\end{tabular}
\end{table*}

\subsection{Baseline Models}
We evaluate our method against several baseline models, including traditional heuristics like Common Neighbors (CN)\cite{Barabasi:1999}, Resource Allocation (RA)\cite{Zhou2009}, and Adamic/Adar (AA)\cite{Adamic:2003}. Additionally, we compare with learning-based models such as Graph Autoencoder (GAE)\cite{Kipf:2016}, Graph Convolutional Networks (GCN)\cite{DBLP:conf/iclr/KipfW17}, and GraphSAGE\cite{Hamilton:2017tp}. Hybrid approaches like Neo-GNN~\cite{yun2021neognns} and BUDDY~\cite{Chamberlain2022GraphNN} are also included, alongside NCNC~\cite{ncnc2024} — the current state-of-the-art in link prediction. Results for baselines are sourced from~\cite{ncnc2024,li2023evaluatinggraphneuralnetworks}.

\subsection{Evaluation Metrics}

We evaluate our model using two standard link prediction metrics: \textbf{Mean Reciprocal Rank (MRR)} and \textbf{Hit Rate at K (HR@K)}. MRR assesses ranking quality by averaging the inverse rank of true positive edges:
\[
\text{MRR} = \frac{1}{N} \sum_{i=1}^{N} \frac{1}{\text{rank}_i}
\]

HR@K measures recall by calculating the fraction of true positives found within the top-K predictions:

\[
\text{HR@K} = \frac{1}{N} \sum_{i=1}^{N} \mathbf{1}\!\left[\text{rank}_i \leq K\right]
\]

where $N$ is the number of test instances, $\text{rank}_i$ is the rank of the true edge for the $i^{th}$ instance, and $\mathbf{1}[\cdot]$ is the indicator function.

To evaluate link prediction performance, we report a single, specific metric tailored to each dataset. The sole metric reported for the FB Page-Page, Coauthor-Physics, Roman-empire, Amazon-ratings, and Chameleon datasets is Mean Reciprocal Rank (MRR). For the DBLP dataset, we exclusively report Hits at 10 (HR@10). All remaining datasets—namely Cora, Citeseer, Pubmed, Facebook, Questions, and Actor—are evaluated using only Hits at 100 (HR@100).

\begin{figure*}[htbp] % Use figure* to span the full page width
    \centering
    
    % --- ROW 1 (3 Heatmaps) ---
    % Width is set to ~0.32\linewidth to fit three images across the page.
    \subfloat[Cora]{\includegraphics[width=0.32\linewidth]{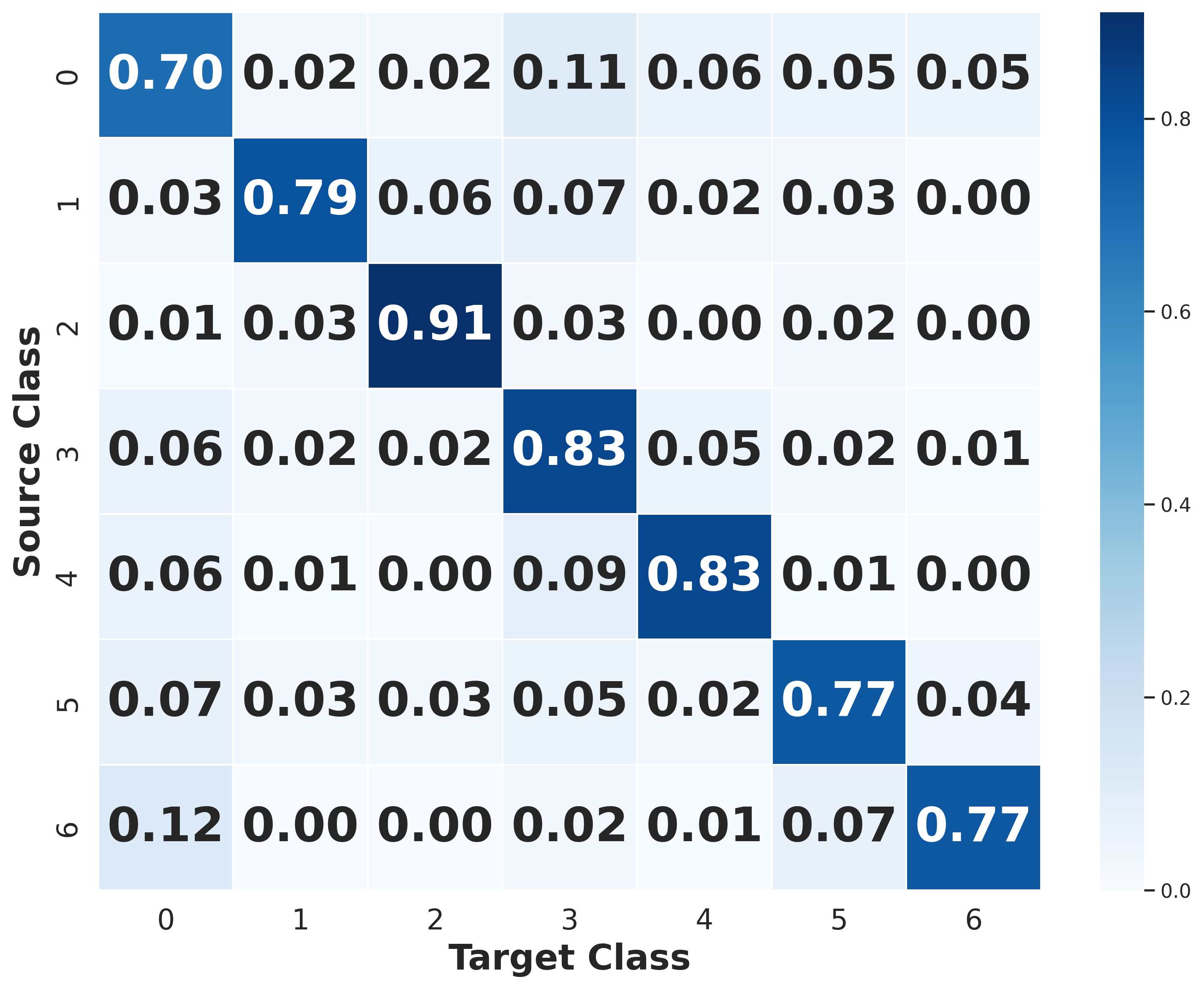}\label{fig:heatmap_Cora}}%
    \hfill
    \subfloat[Citeseer]{\includegraphics[width=0.32\linewidth]{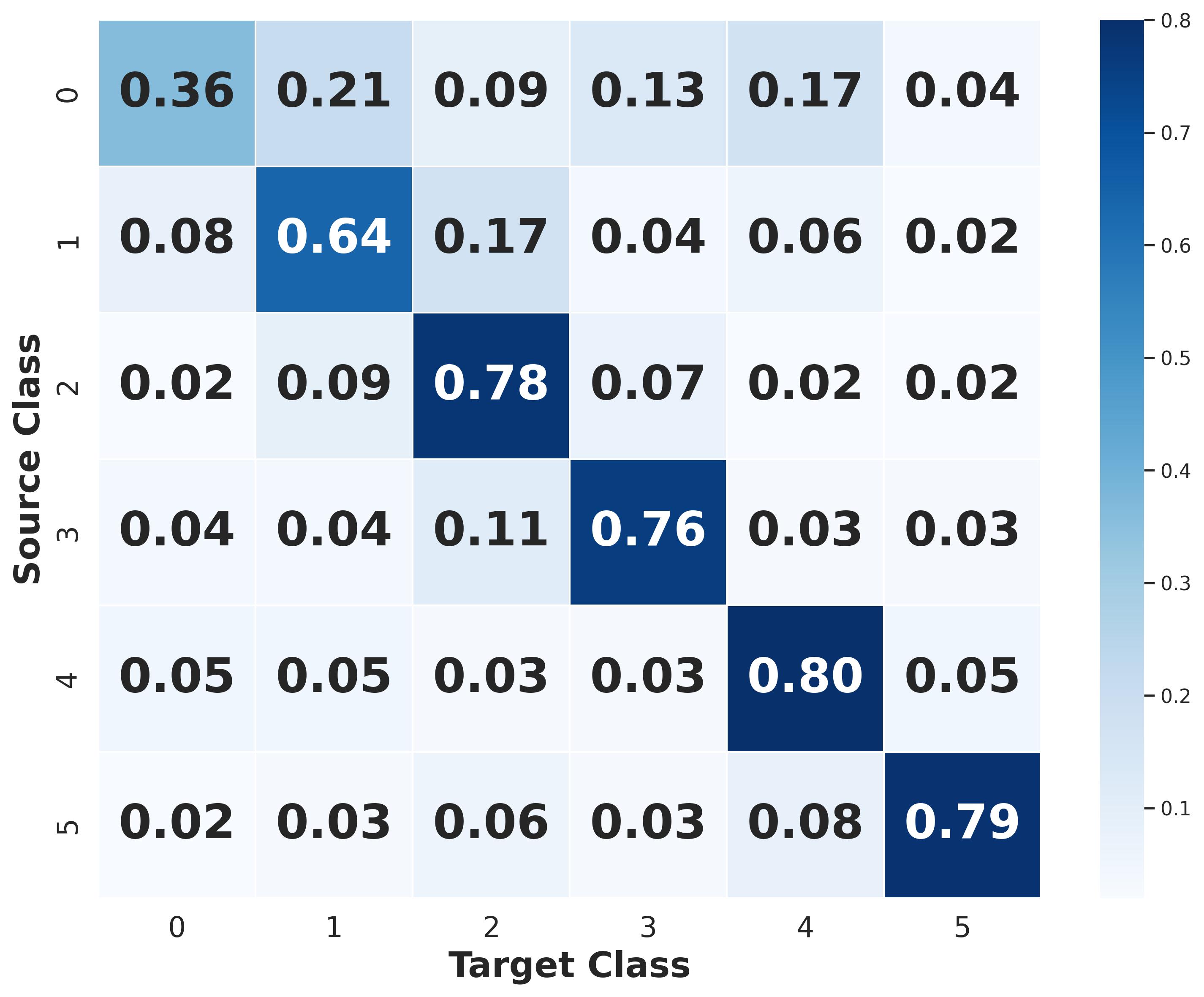}\label{fig:heatmap_Citeseer}}%
    \hfill
    \subfloat[Pubmed]{\includegraphics[width=0.32\linewidth]{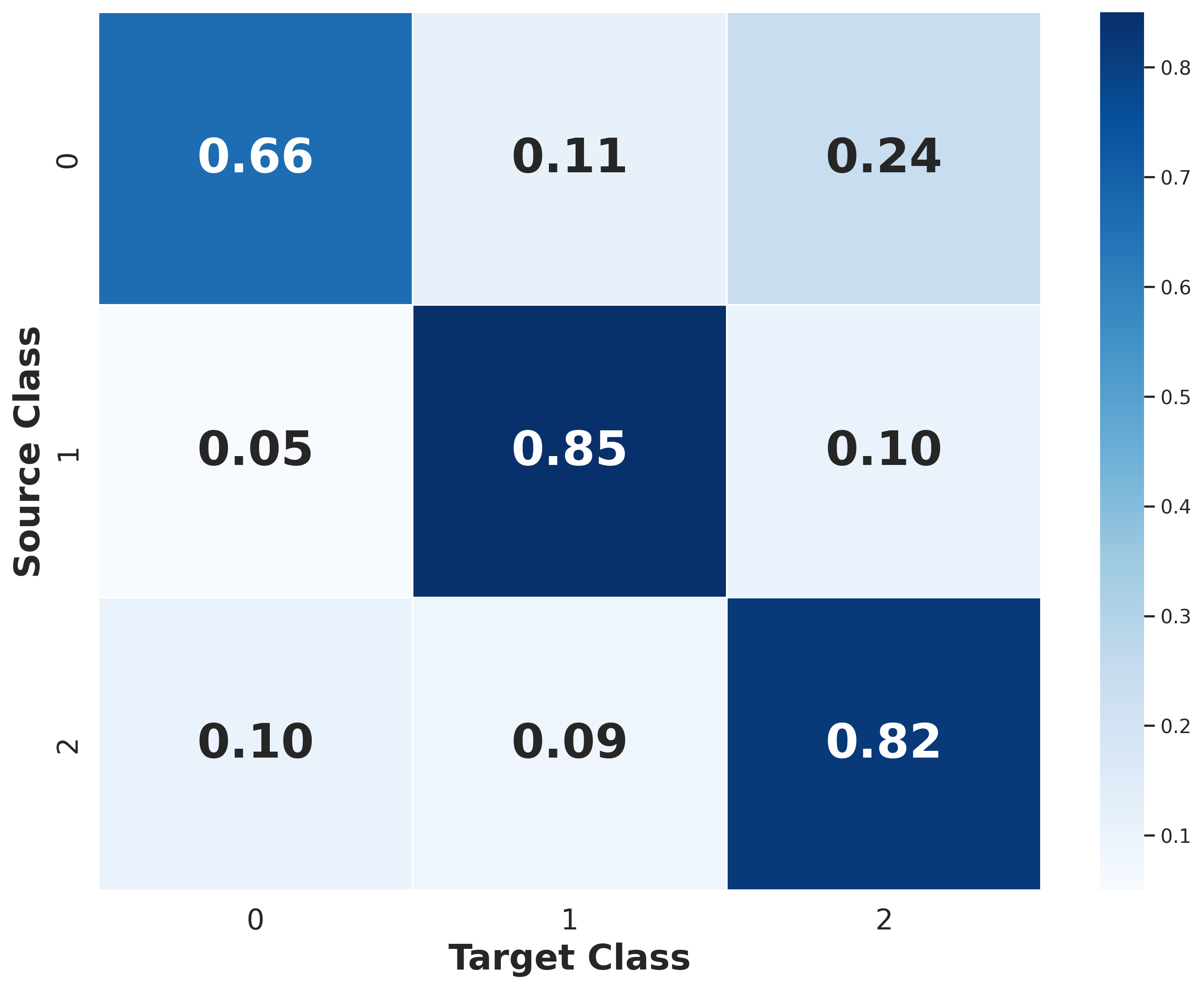}\label{fig:heatmap_Pubmed}}%
    
    \vspace{0.2cm} % Adds a little vertical space between rows

    % --- ROW 2 (3 Heatmaps) ---
    \subfloat[DBLP]{\includegraphics[width=0.32\linewidth]{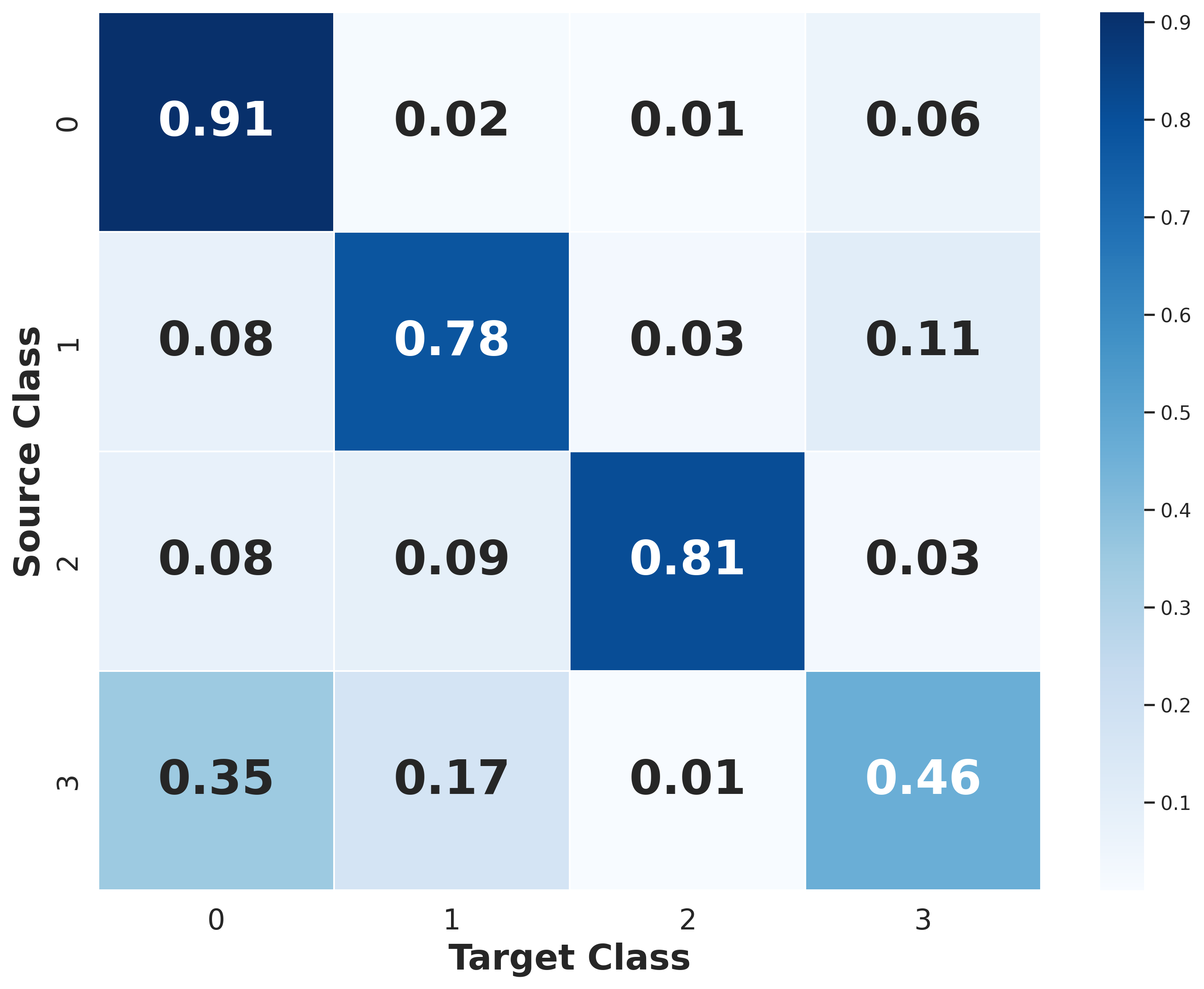}\label{fig:heatmap_DBLP}}%
    \hfill
    \subfloat[Coauthor-Physics]{\includegraphics[width=0.32\linewidth]{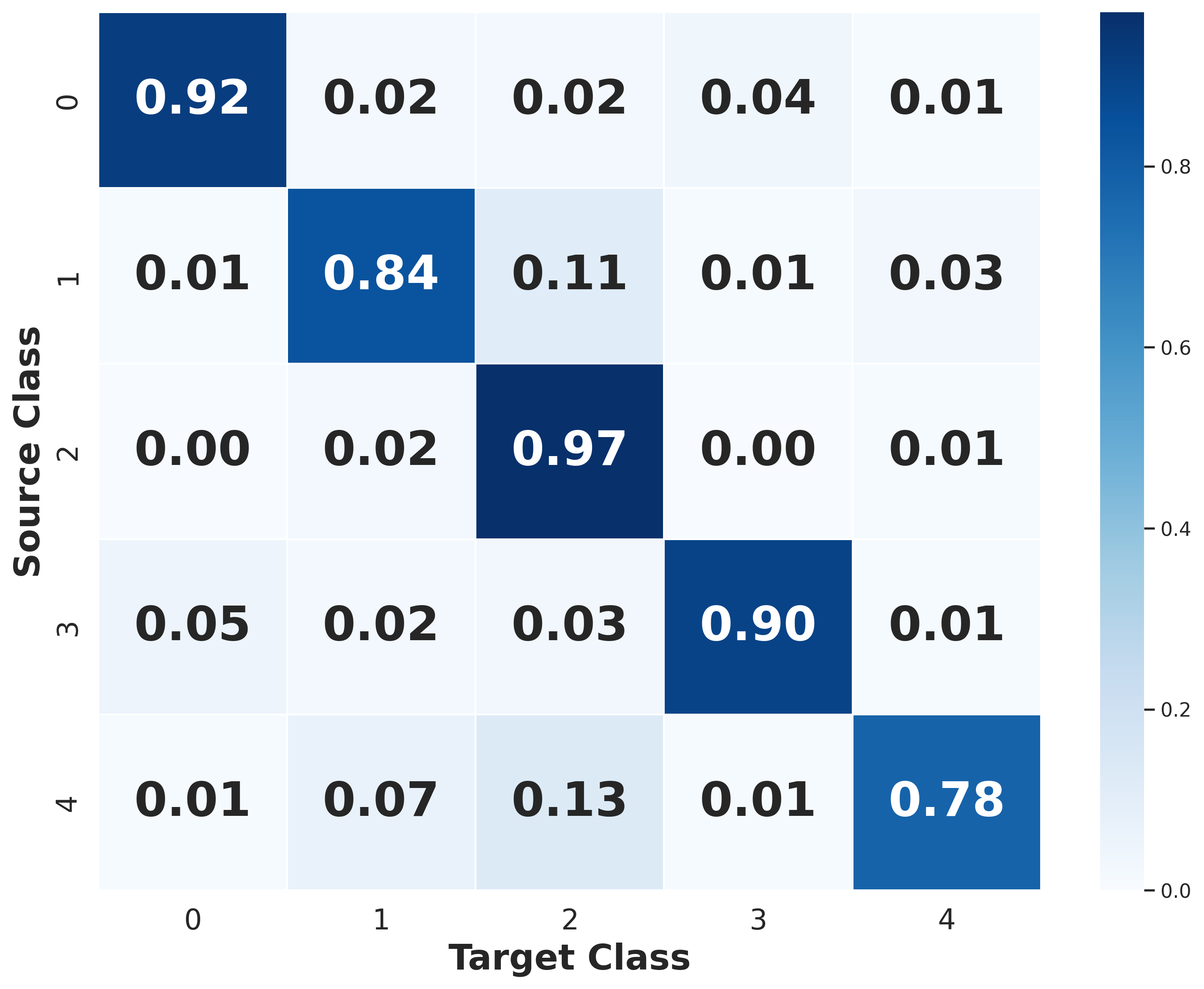}\label{fig:heatmap_cophysics}}%
    \hfill
    \subfloat[Amazon-ratings]{\includegraphics[width=0.32\linewidth]{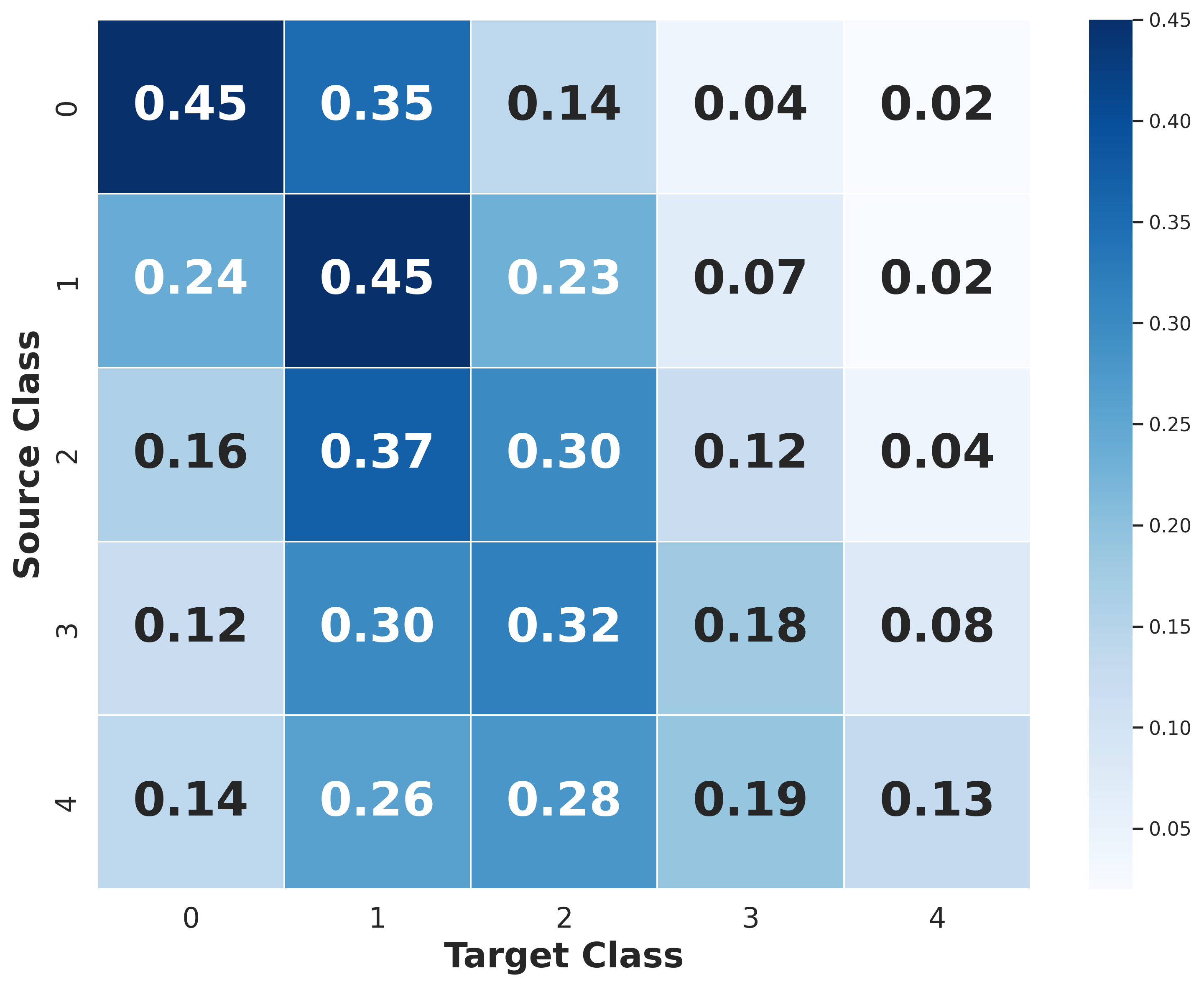}\label{fig:heatmap_amazon}}%
    
    \caption{Heatmaps showing the imbalanced class-class link probabilities across six datasets. This structural imbalance presents a significant challenge for link prediction models, often leading to suboptimal performance.}
    \label{fig:heatmap_analysis_2x3}
\end{figure*}

\section{Ablation Study}
\label{sec:ablation}

In addition to the k-means clustering results presented in Fig.~\ref{fig:k_means}, we implemented and evaluated two alternative methods: Louvain and spectral clustering. For the spectral clustering experiments, we report the performance achieved with the optimal number of clusters, $k$. 

Furthermore, we conducted a mono-label experiment, where all nodes were assigned an identical class label, to evaluate performance without diverse class information. The comprehensive results for all ablation studies are detailed in Table~\ref{table:tab 4} and Table~\ref{table:tab 5}.

\section{{Challenges and Future Directions}}
\label{sec: challenges}

Our analysis reveals that class imbalance presents a significant hurdle. The disproportionate link ratios within datasets like Cora, Citeseer, Pubmed, DBLP, Coauthor-physics, and Amazon-ratings skew the probability matrix $P$ as shown in fig.~\ref{fig:heatmap_analysis_2x3}. This, in turn, compromises model generalization and may result in a minor performance decrease.

Future work should focus on extending the model to \textbf{dynamic} and \textbf{multi-relational graphs} to capture temporal patterns and complex edge types. We also suggest exploring targeted solutions for imbalance, such as specialized loss functions or advanced sampling techniques, and incorporating attention mechanisms to better process node attributes.

% \section{Noise label(Need to be completed)}
% Model performance on four different datasets with varying levels of noise applied to the node features. The plots show the robustness of each model as noise increases.
% \begin{figure}[h!]
% \centering

% % First row of plots
% \begin{subfigure}[b]{0.48\linewidth}
%     \centering
%     \includegraphics[width=\textwidth]{Paper_figures/noise_label_plot/Cora_plot.png}
%     \caption{Cora}
%     \label{fig:Cora_noise}
% \end{subfigure}
% \hfill 
% \begin{subfigure}[b]{0.48\linewidth}
%     \centering
%     \includegraphics[width=\textwidth]{Paper_figures/noise_label_plot/Pubmed_plot.png}
%     \caption{Pubmed}
%     \label{fig:Pubmed_noise}
% \end{subfigure}

% \vspace{0.5cm} % Adds vertical space between the rows

% % Second row of plots
% \begin{subfigure}[b]{0.48\linewidth}
%     \centering
%     \includegraphics[width=\textwidth]{Paper_figures/noise_label_plot/Coauthor-Physics_plot.png}
%     \caption{Coauthor-Physics}
%     \label{fig:cophysics_noise}
% \end{subfigure}
% \hfill
% \begin{subfigure}[b]{0.48\linewidth}
%     \centering
%     \includegraphics[width=\textwidth]{Paper_figures/noise_label_plot/Questions_plot.png}
%     \caption{Questions}
%     \label{fig:Questions_noise}
% \end{subfigure}

% \caption{Model performance on four different datasets with varying levels of noise applied to the node features. The plots show the robustness of each model as noise increases.}
% \label{fig:noise_analysis}

% \end{figure}
\section{Conclusion}
\label{sec: conclude}

In this paper, we introduced \textbf{CGLE}, a framework that successfully augments GNN-based link prediction by incorporating class-level information for both homophilous and heterophilous graphs. Our approach yields superior performance compared to strong models such as NCN and NCNC~\cite{ncnc2024}. We further demonstrated the versatility of CGLE in scenarios without true labels, showing that k-means clustering can generate effective pseudo-labels, with similar success from Louvain and spectral clustering methods. An analysis of the mono-labeling condition also presented. Ultimately, CGLE's blend of accuracy, efficiency, and flexibility makes it a robust and practical tool for graph link analysis applications.

\bibliographystyle{IEEEtran}
\bibliography{reference}

@article{katz1953new,
  author      = {Katz, Leo},
  title       = {A new status index derived from sociometric analysis},
  publication = {Psychometrika},
  year        = {1953},
  url         = {}
}

@inproceedings{brin1998anatomy,
  author      = {Brin, Sergey and Page, Lawrence},
  title       = {The anatomy of a large-scale hypertextual Web search engine},
  publication = {Computer networks and ISDN systems},
  year        = {1998},
  url         = {}
}

@inproceedings{jeh2002simrank,
  author      = {Jeh, Glen and Widom, Jennifer},
  title       = {SimRank: a measure of structural-context similarity},
  publication = {KDD},
  year        = {2002},
  url         = {}
}

@inproceedings{Velickovic:2018we,
  author      = {Veli{\v c}kovi{\'c}, Petar and Cucurull, Guillem and Casanova, Arantxa and Romero, Adriana and Li{\`o}, Pietro and Bengio, Yoshua},
  title       = {{Graph Attention Networks}},
  publication = {ICLR},
  year        = {2018},
  url         = {}
}

@inproceedings{Hamilton:2017tp,
  author      = {Hamilton, William L. and Ying, Zhitao and Leskovec, Jure},
  title       = {{Inductive Representation Learning on Large Graphs}},
  publication = {NIPS},
  year        = {2017},
  url         = {}
}

@inproceedings{ncnc2024,
  author      = {Xiyuan Wang and Haotong Yang and Muhan Zhang},
  title       = {Neural Common Neighbor with Completion for Link Prediction},
  publication = {ICLR},
  year        = {2024},
  url         = {}
}

@inproceedings{zhang2018link,
  author      = {Zhang, Muhan and Chen, Yixin},
  title       = {Link prediction based on graph neural networks},
  publication = {NeurIPS},
  year        = {2018},
  url         = {}
}

@inproceedings{yun2021neognns,
  author      = {Seongjun Yun and Seoyoon Kim and Junhyun Lee and Jaewoo Kang and Hyunwoo J. Kim},
  title       = {Neo-{GNN}s: Neighborhood Overlap-aware Graph Neural Networks for Link Prediction},
  publication = {NeurIPS},
  year        = {2021},
  url         = {}
}

@inproceedings{chamberlain2023graph,
  author      = {Chamberlain, Benjamin Paul and Shirobokov, Sergey and Rossi, Emanuele and Frasca, Fabrizio and Markovich, Thomas and Hammerla, Nils and Bronstein, Michael M and Hansmire, Max},
  title       = {Graph Neural Networks for Link Prediction with Subgraph Sketching},
  publication = {ICLR},
  year        = {2023},
  url         = {}
}

@article{Adamic:2003,
  author      = {Lada A Adamic and Eytan Adar},
  title       = {{Friends and neighbors on the web}},
  publication = {Social networks},
  year        = {2003},
  url         = {}
}

@article{Barabasi:1999,
  author      = {Albert-László Barabási and Réka Albert},
  title       = {{Emergence of scaling in random networks}},
  publication = {Science},
  year        = {1999},
  url         = {}
}

@misc{rozemberczki2021multiscaleattributednodeembedding,
  author      = {Benedek Rozemberczki and Carl Allen and Rik Sarkar},
  title       = {Multi-scale Attributed Node Embedding},
  publication = {arXiv},
  year        = {2021},
  url         = {https://arxiv.org/abs/1909.13021}
}

@misc{shchur2019pitfallsgraphneuralnetwork,
  author      = {Oleksandr Shchur and Maximilian Mumme and Aleksandar Bojchevski and Stephan Günnemann},
  title       = {Pitfalls of Graph Neural Network Evaluation},
  publication = {arXiv},
  year        = {2019},
  url         = {https://arxiv.org/abs/1811.05868}
}

@misc{bojchevski2018deepgaussianembeddinggraphs,
  author      = {Aleksandar Bojchevski and Stephan Günnemann},
  title       = {Deep Gaussian Embedding of Graphs: Unsupervised Inductive Learning via Ranking},
  publication = {arXiv},
  year        = {2018},
  url         = {https://arxiv.org/abs/1707.03815}
}

@article{Yang2020PANESA,
  author      = {Renchi Yang and Jieming Shi and X. Xiao and Yin David Yang and Sourav S. Bhowmick and Juncheng Liu},
  title       = {PANE: scalable and effective attributed network embedding},
  publication = {The VLDB Journal},
  year        = {2020},
  url         = {}
}

@article{Yang2016RevisitingSL,
  author      = {Zhilin Yang and William W. Cohen and Ruslan Salakhutdinov},
  title       = {Revisiting Semi-Supervised Learning with Graph Embeddings},
  publication = {ArXiv},
  year        = {2016},
  url         = {}
}

@inproceedings{Grover:2016,
  author      = {Aditya Grover and Jure Leskovec},
  title       = {{node2vec: Scalable feature learning for networks}},
  publication = {KDD},
  year        = {2016},
  url         = {}
}

@inproceedings{Zhu2021NeuralBN,
  author      = {Zhaocheng Zhu and Zuobai Zhang and Louis-Pascal Xhonneux and Jian Tang},
  title       = {Neural Bellman-Ford Networks: A General Graph Neural Network Framework for Link Prediction},
  publication = {NeurIPS},
  year        = {2021},
  url         = {}
}

@article{Chamberlain2022GraphNN,
  author      = {Benjamin Paul Chamberlain and Sergey Shirobokov and Emanuele Rossi and Fabrizio Frasca and Thomas Markovich and Nils Y. Hammerla and Michael M. Bronstein and Max Hansmire},
  title       = {Graph Neural Networks for Link Prediction with Subgraph Sketching},
  publication = {ArXiv},
  year        = {2022},
  url         = {https://arxiv.org/abs/2209.15486}
}

@article{Zhou2009,
  author      = {Tao Zhou and Linyuan Lü and Yi-Cheng Zhang},
  title       = {Predicting missing links via local information},
  publication = {The European Physical Journal B},
  year        = {2009},
  url         = {}
}

@misc{Kipf:2016,
  author      = {Thomas N. Kipf and Max Welling},
  title       = {Variational Graph Auto-Encoders},
  publication = {arXiv},
  year        = {2016},
  url         = {https://arxiv.org/abs/1611.07308}
}

@inproceedings{Liben-Nowell:2003,
  author      = {David Liben-Nowell and Jon Kleinberg},
  title       = {{The link prediction problem for social networks}},
  publication = {CIKM},
  year        = {2003},
  url         = {}
}

@inproceedings{DBLP:conf/kdd/TylendaAB09,
  author      = {Tomasz Tylenda and Ralitsa Angelova and Srikanta J. Bedathur},
  title       = {Towards time-aware link prediction in evolving social networks},
  publication = {SNAKDD},
  year        = {2009},
  url         = {https://doi.org/10.1145/1731011.1731020}
}

@inproceedings{DBLP:conf/iclr/KipfW17,
  author      = {Thomas N. Kipf and Max Welling},
  title       = {Semi-Supervised Classification with Graph Convolutional Networks},
  publication = {ICLR},
  year        = {2017},
  url         = {}
}

@article{Platonov2023ACL,
  author      = {Oleg Platonov and Denis Kuznedelev and Michael Diskin and Artem Babenko and Liudmila Prokhorenkova},
  title       = {A critical look at the evaluation of GNNs under heterophily: are we really making progress?},
  publication = {ArXiv},
  year        = {2023},
  url         = {https://arxiv.org/abs/2302.11640}
}

@inproceedings{gilmer2017neural,
  author      = {Justin Gilmer and Samuel S. Schoenholz and Patrick F. Riley and Oriol Vinyals and George E. Dahl},
  title       = {Neural Message Passing for Quantum Chemistry},
  publication = {ICML},
  year        = {2017},
  url         = {}
}

@inproceedings{10.1145/3543507.3583340,
  author      = {Song, Xiran and Lian, Jianxun and Huang, Hong and Luo, Zihan and Zhou, Wei and Lin, Xue and Wu, Mingqi and Li, Chaozhuo and Xie, Xing and Jin, Hai},
  title       = {xGCN: An Extreme Graph Convolutional Network for Large-scale Social Link Prediction},
  publication = {WWW},
  year        = {2023},
  url         = {https://doi.org/10.1145/3543507.3583340}
}

@inproceedings{DBLP:conf/www/LeskovecHK10,
  author      = {Jure Leskovec and Daniel P. Huttenlocher and Jon M. Kleinberg},
  title       = {Predicting positive and negative links in online social networks},
  publication = {WWW},
  year        = {2010},
  url         = {https://doi.org/10.1145/1772690.1772756}
}

@article{lloyd1982least,
  author      = {Lloyd, Stuart P.},
  title       = {Least squares quantization in PCM},
  publication = {IEEE Transactions on Information Theory},
  year        = {1982},
  url         = {}
}

@article{rumelhart1986learning,
  author      = {Rumelhart, David E. and Hinton, Geoffrey E. and Williams, Ronald J.},
  title       = {Learning representations by back-propagating errors},
  publication = {Nature},
  year        = {1986},
  url         = {}
}

@article{wang2022equivariant,
  author      = {Wang, Haorui and Yin, Haoteng and Zhang, Muhan and Li, Pan},
  title       = {Equivariant and stable positional encoding for more powerful graph neural networks},
  publication = {arXiv preprint arXiv:2203.00199},
  year        = {2022},
  url         = {https://arxiv.org/abs/2203.00199}
}

@article{koren2009matrix,
  author      = {Koren, Yehuda and Bell, Robert and Volinsky, Chris},
  title       = {Matrix factorization techniques for recommender systems},
  publication = {Computer},
  year        = {2009},
  url         = {}
}

@article{Martnez2016ASO,
  author      = {V{\'i}ctor Mart{\'i}nez and Fernando Berzal and Juan C. Cubero},
  title       = {A Survey of Link Prediction in Complex Networks},
  publication = {ACM Computing Surveys (CSUR)},
  year        = {2016},
  url         = {}
}

@article{Kovcs2018NetworkbasedPO,
  author      = {Istv{\'a}n A. Kov{\'a}cs and Katja Luck and Kerstin Spirohn and Yang Wang and Carl Pollis and Sadie Schlabach and Wenting Bian and Dae-Kyum Kim and Nishka Kishore and Tong Hao and Michael A. Calderwood and Marc Vidal and Albert-Ĺaszl{\'o} Barab{\'a}si},
  title       = {Network-based prediction of protein interactions},
  publication = {Nature Communications},
  year        = {2018},
  url         = {}
}

@article{Yue2019GraphEO,
  author      = {Xiang Yue and Zhen Wang and Jingong Huang and Srinivasan Parthasarathy and Soheil Moosavinasab and Yungui Huang and S. Lin and Wen Zhang and Ping Zhang and Huan Sun},
  title       = {Graph embedding on biomedical networks: methods, applications, and evaluations},
  publication = {Bioinformatics},
  year        = {2019},
  url         = {}
}

@book{DBLP:books/daglib/0025903,
  author      = {David A. Easley and Jon M. Kleinberg},
  title       = {Networks, Crowds, and Markets - Reasoning About a Highly Connected World},
  publication = {Cambridge University Press},
  year        = {2010},
  url         = {https://doi.org/10.1017/CBO9780511761942}
}

@inproceedings{Pei2020Geom-GCN:,
  author      = {Hongbin Pei and Bingzhe Wei and Kevin Chen-Chuan Chang and Yu Lei and Bo Yang},
  title       = {Geom-GCN: Geometric Graph Convolutional Networks},
  publication = {ICLR},
  year        = {2020},
  url         = {}
}

@inproceedings{DBLP:conf/aaai/ShiHZHZZ24,
  author      = {Lei Shi and Bin Hu and Deng Zhao and Jianshan He and Zhiqiang Zhang and Jun Zhou},
  title       = {Structural Information Enhanced Graph Representation for Link Prediction},
  publication = {AAAI},
  year        = {2024},
  url         = {https://doi.org/10.1609/aaai.v38i13.29417}
}

@article{DBLP:journals/pnas/BensonASJK18,
  author      = {Austin R. Benson and Rediet Abebe and Michael T. Schaub and Ali Jadbabaie and Jon M. Kleinberg},
  title       = {Simplicial closure and higher-order link prediction},
  publication = {Proc. Natl. Acad. Sci. {USA}},
  year        = {2018},
  url         = {}
}

@article{Battiston2020NetworksBP,
  author      = {Federico Battiston and Giulia Cencetti and Iacopo Iacopini and Vito Latora and Maxime Lucas and Alice Patania and Jean-Gabriel Young and Giovanni Petri},
  title       = {Networks beyond pairwise interactions: structure and dynamics},
  publication = {ArXiv},
  year        = {2020},
  url         = {https://api.semanticscholor.org/CorpusID:219179840}
}

@inproceedings{DBLP:conf/pkdd/AbuodaMA19,
  author      = {Ghadeer Abuoda and Gianmarco De Francisci Morales and Ashraf Aboulnaga},
  title       = {Link Prediction via Higher-Order Motif Features},
  publication = {ECML PKDD},
  year        = {2019},
  url         = {}
}

@misc{li2023evaluatinggraphneuralnetworks,
  author      = {Juanhui Li and Harry Shomer and Haitao Mao and Shenglai Zeng and Yao Ma and Neil Shah and Jiliang Tang and Dawei Yin},
  title       = {Evaluating Graph Neural Networks for Link Prediction: Current Pitfalls and New Benchmarking},
  publication = {arXiv},
  year        = {2023},
  url         = {https://arxiv.org/abs/2306.10453}
}

@INPROCEEDINGS{6268901,
  author={Sulaiman, Nor Hashimah and Mohamad, Daud},
  booktitle={2012 IEEE Symposium on Humanities, Science and Engineering Research}, 
  title={A Jaccard-based similarity measure for soft sets}, 
  year={2012},
  doi={10.1109/SHUSER.2012.6268901}}

\end{document}